\documentclass[fleqn,usenatbib]{mnras}

\usepackage{duckuments}

\usepackage{newtxtext,newtxmath}
\usepackage[T1]{fontenc}
\usepackage{graphicx}

\usepackage{amsmath,gensymb,mathrsfs}
\usepackage{booktabs}
\usepackage{rotating}
\usepackage{pdflscape}
\usepackage{CJK}
\usepackage{lineno}

\newcommand{\new}[1]{#1}

\newcommand{\NSDB}{$30$}

\title[Asteroseismic hot subdwarf magnetometry]{Tight asteroseismic upper limits on hot subdwarf magnetism}

\author[N. Z. Rui \& L. Ma]{
Nicholas Z. Rui (芮峥嵘),$^{1,2}$\thanks{E-mail: nrui@princeton.edu, NASA Hubble Fellow}\label{emails}\hyperref[contrib]{\footnotemark[3]}
and Linhao Ma (马林昊)$^{1,3}$\thanks{E-mail: linhaoma@princeton.edu}\thanks{These authors contributed equally to this work.}\label{contrib}
\\
$^{1}$Department of Astrophysical Sciences, Princeton University, 4 Ivy Lane, Princeton, NJ 08544, USA\\
$^{2}$Center for Interdisciplinary Exploration and Research in Astrophysics (CIERA), Northwestern University, 1800 Sherman Ave., Evanston, IL 60201, USA\\
$^{3}$Kavli Institute for Theoretical Physics, University of California, Santa Barbara, CA 93106, USA
}

\date{Accepted XXX. Received YYY; in original form ZZZ}
\pubyear{\the\year{}}

\begin{document}
\begin{CJK*}{UTF8}{gbsn}
\label{firstpage}
\pagerange{\pageref{firstpage}--\pageref{lastpage}}
\maketitle

\begin{abstract}
The apparent absence of magnetism in subdwarf B stars (sdBs) has emerged as an outstanding observational puzzle, especially in light of their likely origin in binary interactions.
Gravity-mode pulsations in some sdBs offer a unique opportunity to extend recent successes in asteroseismic magnetic field detection in red giants.
In the presence of strong magnetic fields, gravity modes are expected to be significantly damped, suppressing their observed amplitudes.
We place stringent upper limits on magnetic fields in 30 hot subdwarfs by using the observation of low-frequency dipole and quadrupole modes to rule out magnetic suppression.
Even under conservative assumptions, our upper limits constrain near-surface magnetic fields to below a kilogauss in roughly half of our sample, with the strongest individual constraint reaching down to $\sim20\,\mathrm{G}$.
These upper limits reinforce the apparent dearth of magnetic fields in hot subdwarfs and demonstrate asteroseismology as a powerful new probe of magnetism in these stars.
\end{abstract}


% \linenumbers

\section{Introduction} 

\begin{figure*}
    \centering
    \includegraphics[width=\linewidth]{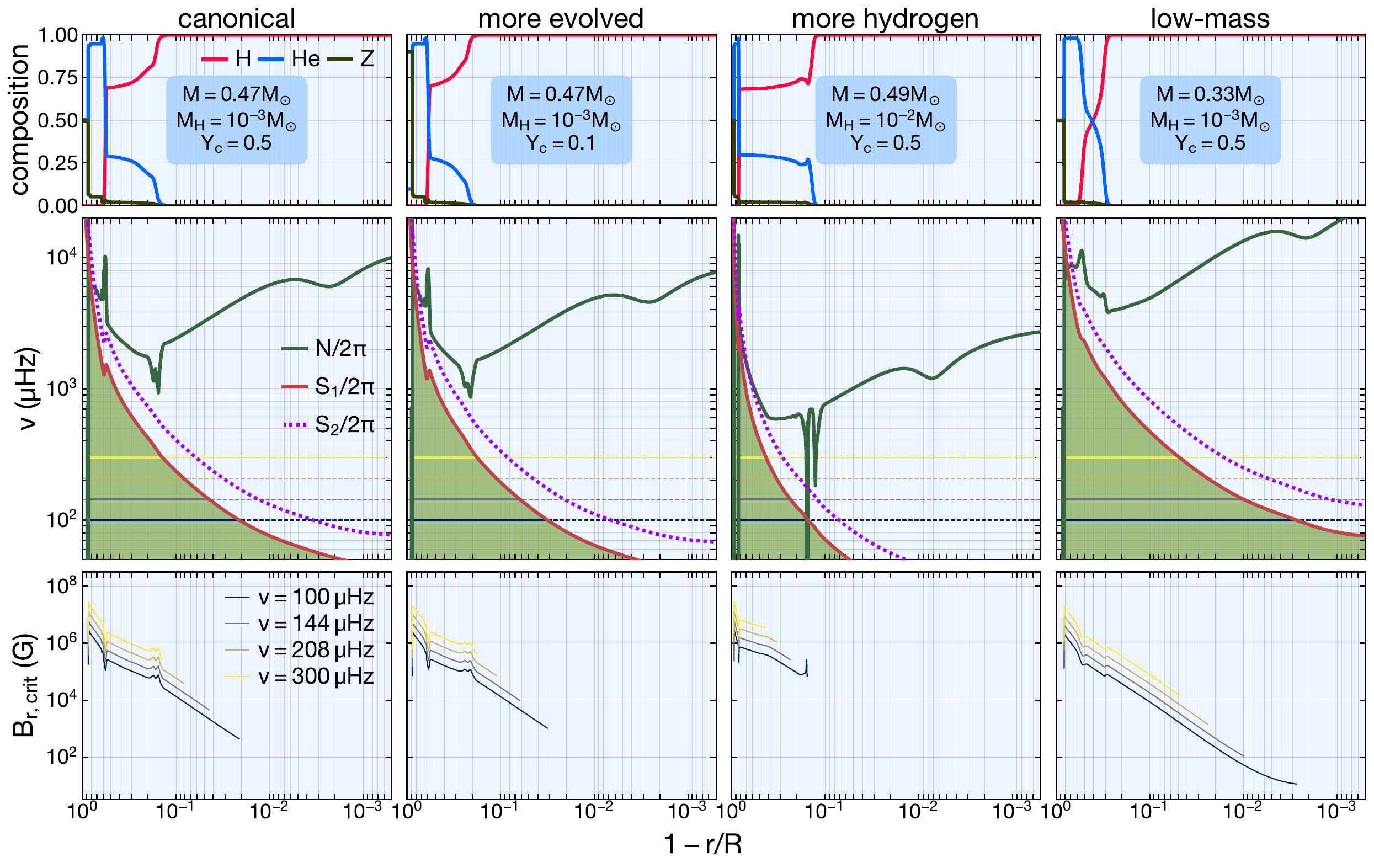}
    \caption{Four sdB stellar models with varying parameters.
    \textit{First row:} Composition profile.
    \textit{Second row:} Asteroseismic propagation diagram, showing the Brunt--V\"ais\"al\"a frequency $N$ and dipole ($\ell=1$) and quadrupole ($\ell=2$) Lamb frequencies $S_1$ and $S_2$.
    The shaded region corresponds to the g-mode cavity for dipole modes.
    Horizontal lines sample the range of periods exhibited by g-mode pulsations in observed sdBs.
    \textit{Third row:} For various mode periods, the critical magnetic field $B_{r,\mathrm{crit}}$ at which g-mode suppression is expected (Equation \ref{eqn:bcrit}).}
    \label{fig:FIG_compare_models}
\end{figure*}

B-type hot subdwarf stars (sdBs) are core helium burning stars with masses $\lesssim0.5M_\odot$.
In the field, sdBs are predominantly thought to form through binary interactions, either in a merger of two helium white dwarfs \citep{Zhang:2012:DoubleHeWDMergers,Schwab:2018:HeWDMerger} or in the removal of a red giant's envelope during either a common-envelope event \citep{Xiong:2017:HotSubdwarfCEE} or a phase of stable mass transfer \citep{Vos:2020:sdBSMT}.
Each of these channels is expected to contribute significantly to the total formation rate of sdBs \citep{Han:2002:sdBFormation,Han:2003:sdBFormation,Dawson:2026:sdBdata}.
Stars resulting from mergers are expected to be single \citep[e.g.,][]{Heber:2009:sdBReview}, whereas those resulting from envelope stripping are expected to be in binaries \citep{Maxted:2001:sdBBinarity}.

Although there are thousands of spectroscopically confirmed hot subdwarfs \citep{Culpan:2022:HotSubdwarfCatalog}, only $7$ have been confirmed to be magnetic, with large field strengths $180$--$450\,\mathrm{kG}$ \citep[][although see \citealt{Jeffery:2026:SALTHotSubdwarfs} and \citealt{Vos:2021:PostMergerSubdwarf}]{Dorsch:2022:MagneticsdO,Pelisoli:2022:MagneticsdOs,Dorsch:2024:MagneticsdOs,Jeffery:2026:SALTHotSubdwarfs}.
None of these objects are sdB stars, instead all belonging to an exceptionally rare class of higher-mass ($M\simeq0.8M_\odot$, $T_{\mathrm{eff}}\gtrsim45\,\mathrm{kK}$) O-type hot subdwarfs with intermediate helium-rich compositions and a mysterious spectroscopic feature at 4630\,\AA.
These so-called intermediate helium-rich sdO-type (iHe-sdO) stars are very likely to be merger remnants \citep[e.g.,][]{Justham:2011:HerichSubdwarfFormation}.
Spectropolarimetric searches for magnetic fields in around two dozen hot subdwarfs (of both sdB- and sdO-type) have ruled out magnetic fields $\gtrsim1$--$2\,\mathrm{kG}$ in all cases \citep[see][and the references therein]{Landstreet:2012:SpectropolSubdwarfs}.
In some cases, the inferred upper limits are as low as a few hundred gauss.
The dearth of detectable magnetic fields in the vast majority of hot subdwarfs is puzzling.
Both white dwarf mergers \citep{Beloborodov:2014:MagneticOutbursts,Schneider:2016:MergerMagnetism,Schneider:2019:MergerMagnetism,Pakmor:2024:HeWDMergers} and common-envelope events \citep{Tout:2008:CEDynamo,Nordhaus:2011:CEEMagnetism} are hydrodynamically complex, strongly shearing flows in which magnetic dynamos are expected to be efficient.
\citet{Pakmor:2024:HeWDMergers} suggest that following the merger of an unequal-mass helium white dwarf binary, an extended period of helium shell burning could destroy large-scale ordered fields generated during the merger \citep[see also][]{Zhang:2012:DoubleHeWDMergers,Schwab:2018:HeWDMerger}.
However, it remains an enormous technical challenge to simulate the evolution of these magnetic fields in the merger remnant over stellar evolutionary timescales.

Although stellar magnetic fields are usually measured using spectroscopy or spectropolarimetry \citep{Donati:2009:Zeeman}, asteroseismology has recently emerged as a powerful method for probing stellar magnetism.
Magnetic fields with strengths $\simeq10\mathrm{s}$--$100\mathrm{s}\,\mathrm{kG}$ have now been asteroseismically measured in the cores of $\approx70$ red giants \citep{Li:2022:30to100kG,Deheuvels:2023:MagneticRG,Li:2023:13Magnetic,Hatt:2024:MagneticRG,Villate:2026:MagneticOffset}.
The detection of magnetic suppression in $\approx20\%$ of red giants implies the presence of even stronger core magnetic fields $\gtrsim100\mathrm{s}\,\mathrm{kG}$ \citep{Fuller:2015:SuppressedDipole,Stello:2016:MagneticPrevalence}.
Both of these techniques exploit the fact that gravity (g) modes can be observably modified \citep{Bugnet:2021:MagneticI,Rui:2024:TARM,Lignieres:2024:MagnetoGravitoinertial} or even significantly damped \citep{Fuller:2015:SuppressedDipole,Lecoanet:2017:MagneticConversion,Rui:2023:MagneticSuppression,David:2025:Magnetogravity} by magnetic fields.
Although, for practical reasons, red giants have been the main subject of asteroseismic magnetometry, the underlying physics is generic and expected to apply to all g-mode pulsations.
Asteroseismic detections of magnetic fields have already been reported in a handful of main-sequence g-mode pulsators \citep[see reported asteroseismic detections of magnetic fields in ][]{Lecoanet:2022:MagneticSPB,Vandersnickt:2025:BcepMagnetic,Takata:2026:ToroidalGDor}.

Asteroseismic upper limits on stellar magnetic fields can also be derived from the null detection of magnetic effects in g modes.
\citet{Rui:2025:wdmagnet} use the absence of magnetic frequency shifts and magnetic suppression to place stringent upper limits on the near-surface magnetic fields of \new{24} pulsating white dwarfs \citep[see also][]{Blatman:2025:ultramagnet}.
These upper limits are typically $\sim1$--$10\,\mathrm{kG}$, but reach as low as $\sim10$--$100\,\mathrm{G}$ for white dwarfs possessing longer-period modes ($P\gtrsim800\,\mathrm{s}$).
\new{For comparison, spectroscopy and spectropolarimetry are sensitive to white dwarf magnetic fields with strengths $\gtrsim1\,\mathrm{MG}$ and $\gtrsim1$--$10\,\mathrm{kG}$, respectively \citep{Donati:2009:Zeeman,Bagnulo:2021:MagneticWDs}.}

Hundreds of sdBs are known to exhibit opacity-mechanism-driven pulsations, both in pressure and gravity modes \citep{Charpinet:1996:PulsatingsdBs,Kilkenny:2007:PulsatingHotSubdwarfs,Uzundag:2024:PulsatingHotSubdwarfs}.
sdBs displaying g-mode pulsations with characteristic periods $\simeq1$--$2\,\mathrm{hr}$ (also called V1093 Her stars) usually possess cooler effective temperatures $\lesssim27\,\mathrm{kK}$, with some objects close to this threshold displaying hybrid pulsations (i.e., both pressure and gravity modes).
Members of a small class of intermediate-helium-rich sdBs have been found to pulsate in g modes at higher effective temperatures $\approx35\,\mathrm{kK}$ \citep[``V366 Aquarid'' stars;][]{Ahmad:2005:sdBData,Latour:2019:sdBData,Oestensen:2020:sdBData}.

In this study, we place asteroseismic upper limits on the near-surface magnetic fields of \NSDB{} g-mode pulsating hot subdwarfs, applying theoretical insights from the parallel study of internal magnetic fields in pulsating red giants.
Our choices are designed to place maximally conservative upper limits on the magnetic field.
In spite of this, we demonstrate that asteroseismology often places extraordinarily strong constraints on sdB magnetism.

\section{Methods}

\subsection{Magnetic suppression of g-mode pulsations}

Gravity waves are buoyantly restored hydrodynamical waves which propagate in stably stratified (radiative) regions.
Standing gravity waves (g modes) oscillate at a discrete set of frequencies.
When precisely measured, these mode frequencies yield valuable information about the interior of the star.
To propagate, g modes must have a frequency $\nu$ lower than both $N/2\pi$ and $S_\ell/2\pi$, where $N$ is the Brunt--V\"ais\"al\"a frequency \citep{Vaisala:1925:BruntVaisala} and $S_\ell=\sqrt{\ell(\ell+1)}c_s/r$ is the Lamb frequency ($\ell$ is the angular degree of the mode and $c_s$ is the speed of sound).
In particular, the Brunt--V\"ais\"al\"a frequency $N$ describes the frequency of buoyantly restored oscillations and measures the degree to which a layer of the star is stratified.
This condition defines a radial region called the g-mode cavity, which we denote as $\mathcal{R}^\ell_{\;\nu}$.
The second row of Figure \ref{fig:FIG_compare_models} shows the g-mode cavity regions for four sdB stellar models with varying parameters (green shaded regions).

The g modes found in sdBs have high radial orders $\sim$ tens, so that roughly $\nu\ll N/2\pi$.
This implies that the associated fluid is strongly confined by stratification, so that most of its motion is horizontal.
Because the magnetic tension (the relevant component of the Maxwell stress) restores motions perpendicular to the magnetic field, g modes are primarily sensitive to the radial component $B_r$ of the field.
In this study, we proceed with the assumption that the horizontal component of the field $B_h$ can be ignored, which is valid as long as $B_h/B_r$ is not too large ($B_h/B_r\lesssim N/\omega$, where $N/\omega\gg 1$).
Correspondingly, this study constrains only $B_r$, which may not represent the true magnetic field strength if the horizontal component $B_h$ is dominant.

When a gravity wave propagates in a strongly magnetized region with $B_r$ larger than
\begin{equation} \label{eqn:bcrit}
    B_{r,\mathrm{crit}} \sim 4\pi^2a_{\mathrm{crit}}\frac{\sqrt{4\pi\rho}\nu^2r}{N}\mathrm{,}
\end{equation}
the Lorentz force becomes comparable to buoyancy \citep{Fuller:2015:SuppressedDipole,Cantiello:2016:EvolvingMagnetic}.
Under these circumstances, the gravity wave is expected to efficiently damp through conversion to slow magnetic \citep{Lecoanet:2017:MagneticConversion,Rui:2023:MagneticSuppression} and Alfv\'en waves \citep{Loi:2017:AlfvenResonances,Lecoanet:2022:MagneticSPB,Rui:2023:MagneticSuppression,David:2025:Magnetogravity}.
The dimensionless prefactor $a_{\mathrm{crit}}$ depends on the quantum numbers $\ell$ and $m$ as well as the geometry of the field, but we follow \citet{Fuller:2015:SuppressedDipole} in roughly estimating it as $a_{\mathrm{crit}}=1/2\sqrt{\ell(\ell+1)}$ \citep[more realistic values of $a_{\mathrm{crit}}$ are only expected to affect our results by a factor of a few;][]{Lecoanet:2022:MagneticSPB,Rui:2023:MagneticSuppression,David:2025:Magnetogravity}.
The third row of Figure \ref{fig:FIG_compare_models} shows the the $B_{r,\mathrm{crit}}$ profile for various sdB stellar models at different realistic mode frequencies.
As can be seen in Equation \ref{eqn:bcrit}, magnetic g-mode suppression occurs at lower magnetic field thresholds for lower-frequency g modes in regions with lower densities $\rho$ and greater $N$.
In our models, g modes are usually most easily suppressed in the outermost (lowest-density) layers of the g-mode cavity \citep[similar to the case in pulsating white dwarfs;][]{Rui:2025:wdmagnet}.
Accurate asteroseismic inference of the magnetic field therefore rests on careful modeling of the outer layers of the stellar structure.
Occasionally the minimum $B_{r,\mathrm{crit}}$ is instead attained by a highly stratified layer at the hydrogen-to-helium interface, in which case a strong constraint on the internal magnetic field is also possible.

\subsection{Models} \label{sec:models}

\begin{figure}
    \centering
    \includegraphics[width=\linewidth]{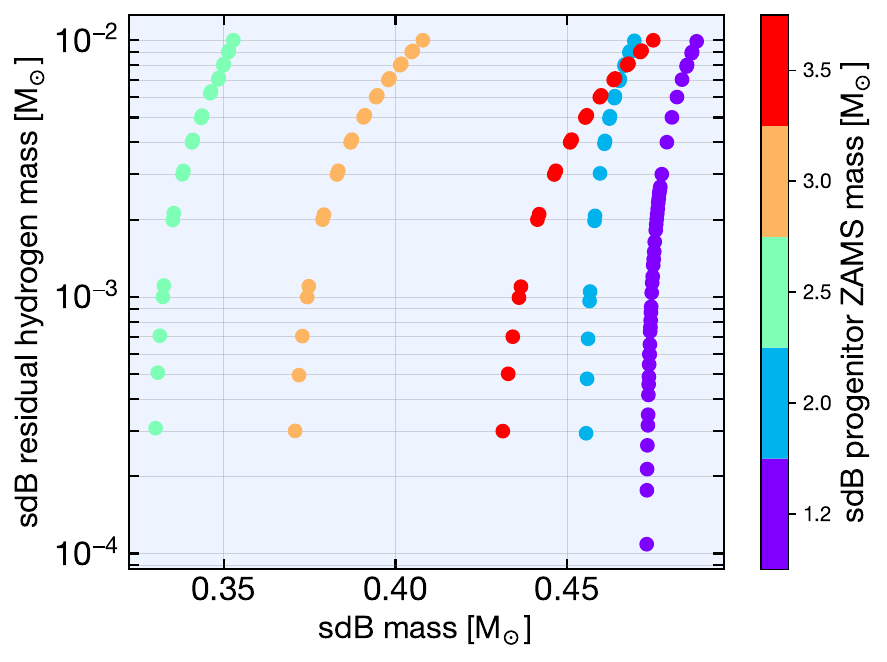}
    \caption{Our grid of sdB \texttt{MESA} models, with each point corresponding to a unique zero-age sdB model.
    We start from 5 different progenitor masses (indicated by the color bar) corresponding to the progenitors of both canonical- and low-mass sdBs.
    From each progenitor model, we create a number of different sdB models with varying amounts of residual hydrogen masses $10^{-4}M_\odot\leq M_{\mathrm{H}}\leq10^{-2}M_\odot$, providing good coverage over the range of observed sdB parameters.
    }
    \label{fig:FIG_sdb_grid}
\end{figure}

The critical magnetic field $B_{r,\mathrm{crit}}$ in Equation \ref{eqn:bcrit} depends on the stellar structure.
In order to evaluate $B_{r,\mathrm{crit}}$ for observed sdBs, we construct one-dimensional sdB evolutionary models using the \texttt{MESA} stellar evolution code (version r24.08.1; \citealt{Paxton:2011:MESA,Paxton:2013:MESA,Paxton:2015:MESA,Paxton:2018:MESA,Paxton:2019:MESA,Jermyn:2023:MESA}).
This Section describes the details of our model setup, which closely follows that of \cite{Ma:2024:sdBTidalSpinup}.

\subsubsection{Zero-age sdB model grid}

Many sdB stars are thought to form through binary interactions that strip nearly the entire hydrogen-rich envelopes of their progenitors shortly before core helium ignition near the tip of the red giant branch, or TRGB (\citealt{Han:2002:sdBFormation,Han:2003:sdBFormation,Pelisoli:2020:sdBFormation,Scherbak:2023:CommonEnvelope}; see also Section \ref{sec:results} for a discussion of the merger channel for forming single sdBs).
This envelope stripping process leaves behind a helium-burning core whose mass is fixed to the core mass of the progenitor.
During this process, the star is also expected to retain a small amount of residual hydrogen.
Instead of explicitly modeling these binary interactions, we construct a grid of sdBs using single-star evolution models and mimic the envelope stripping through artificial mass removal. This approach is more aligned with envelope stripping caused by common envelope ejection, which is the expected scenario for the short-period ($P\lesssim30\,\mathrm{days}$) binaries in our sample \citep{Chen:2013:sdBformation}.

We start from a grid of five sdB progenitor models with zero-age main-sequence (ZAMS) masses $M=1.2$, $2.0$, $2.5$, $3.0$, and $3.5M_\odot$.
For simplicity, all models have solar metallicity ($Z=0.02$).
The $1.2M_\odot$ model represents the progenitors of ``canonical-mass'' sdBs.
The progenitors of these sdBs, typically with ZAMS masses $\lesssim2M_\odot$, develop degenerate helium cores and ignite helium through an off-center helium flash near the TRGB which later develops into core helium burning. Because the helium-core mass at ignition is only weakly dependent on the progenitor mass in this regime, envelope stripping produces sdBs with a characteristic canonical mass of $\approx0.47M_\odot$.
The more massive progenitors in our grid ignite helium without forming a degenerate core, resulting in ``low-mass'' sdBs with $\lesssim0.47M_\odot$.
Observational studies suggest that both canonical sdBs and these lower-mass systems may constitute substantial fractions of the sdB population (e.g., Fig.~11 of \citealt{Schaffenroth:2022:sdBBinaries}).
Our adopted progenitor grid is therefore designed to sample both regimes.

The amount of residual hydrogen in sdBs, despite its tiny mass ($\lesssim10^{-2}M_\odot$), controls the radius and density of the envelope to which our magnetic constraints are most sensitive.
The mass of residual hydrogen is observationally constrained by comparing the spectroscopically measured values of $\log g$ and $T_{\mathrm{eff}}$ to stellar models, with typical values $M_{\mathrm{H}}\sim10^{-3}$--$5\times10^{-3}M_\odot$ (see, e.g., Fig.~10 of \citealt{Kupfer:2015:sdBData}), where larger values of $M_{\mathrm{H}}$ correspond to more inflated envelopes and lower surface gravities.
Therefore, for each of our progenitor models, we construct a variety of zero-age sdB models with different values of $M_{\mathrm{H}}$.
For the $1.2M_\odot$ progenitor, we choose different values of $M_{\mathrm{H}}$ with $10^{-4}M_\odot\leq M_{\mathrm{H}}\leq10^{-2}M_\odot$, while for the other progenitors, we choose hydrogen masses $3\times 10^{-4}M_\odot\leq M_{\mathrm{H}}\leq10^{-2}M_\odot$.

We evolve the progenitor models from the pre-main sequence to the TRGB, which we define as the point at which $T_{\mathrm{eff}}$ reaches its minimum following core hydrogen depletion.
We then use the \texttt{relax\_mass} routine in \texttt{MESA} to remove the outer layers of the progenitor models until the remaining hydrogen mass in the model reaches the desired $M_{\mathrm{H}}$ in our grid.
We take the resulting stripped model as a zero-age sdB and evolve it through core helium burning until its central helium mass fraction $Y_c$ falls below $1\%$.
Although a few models do not numerically converge, we produce a model grid that covers a variety of progenitor masses and residual hydrogen masses (Figure \ref{fig:FIG_sdb_grid}).

\subsubsection{sdB evolution under element diffusion and mixing}

Our estimates of $B_{r,\mathrm{crit}}$ rely on accurate modeling of the sdB envelope, which contains a significant portion of the g-mode cavity.
In order to create realistic chemical profiles of these outer layers, we include detailed elemental diffusion and mixing physics.

We turn on element diffusion using \texttt{MESA} diffusion-class representatives $^1\mathrm{H}$, $^4\mathrm{He}$, $^{12}\mathrm{C}$, $^{14}\mathrm{N}$, and $^{16}\mathrm{O}$ during the sdB core-helium burning phase.
This includes consistent treatments of gravitational settling and radiative acceleration \citep{Paxton:2018:MESA}.
We find that the radiative envelope in our models, where gravity waves propagate, stratifies into a hydrogen-dominated outer envelope and a helium-dominated inner envelope.
This is consistent with the low surface helium abundances measured in sdBs, especially for old systems \citep[e.g.,][]{Heber:2009:sdBReview}.

We apply the ``predictive mixing'' scheme for mixing near the convective core boundaries \citep{Paxton:2018:MESA}.
This causes the convective core in our models to grow steadily over time during the core helium burning phase, consistent with the findings of asteroseismic studies that core helium burning stars tend to possess larger convective cores than theoretically expected \citep{Constantino:2015:LargeConvCore,Bossini:2017:LargeConvCore,Noll:2024:LargeConvCore}.
Due to this agreement, predictive mixing is preferred over other treatments concerning mixing at the convective-zone boundaries, including those reliant on convective overshoot parameters, which are poorly constrained by observations.

Because we do not enforce a minimum diffusion coefficient in our models (e.g., \texttt{min\_D\_mix} in \texttt{MESA}), the Brunt--V\"ais\"al\"a frequency profiles we obtain sometimes develop huge spikes at compositional discontinuities, especially near the boundary of the convective core and the hydrogen--helium interface in the envelope.
Equation \ref{eqn:bcrit} then na\"ively implies extremely low values of $B_{r,\mathrm{crit}}$ in the neighborhood of these discontinuities.
Due to uncertainties in internal stellar mixing processes, our stellar models should not be relied upon to accurately predict the stratification structures in these regions.
Furthermore, our critical magnetic field calculation \new{assumes that the wavelength is short compared to length scales over which the background varies appreciably} 
\citep[e.g.,][]{Lecoanet:2017:MagneticConversion}\new{.
This condition breaks down in the vicinity of these buoyancy glitches.}
Because these spikes usually exist deep inside the stellar envelope, we restrict our $B_{r,\mathrm{crit}}$ evaluation to regions with a hydrogen mass fraction $X\geq10^{-4}$ in our models in order to calculate a maximally conservative $B_{r,\mathrm{crit}}$.
Restricting our analysis to the outer envelope also removes the dependence of our analysis on ``breathing pulses'' which occur near the end of the helium burning phases.
These breathing pulses \citep[which have been argued to be numerical artifacts;][]{Bauer:2021:BreathingPulses} create small transient radiative cavities in the convective cores in which ultra-short wavelength g modes can theoretically propagate.

\subsection{Placing asteroseismic upper limits on magnetic fields} \label{sec:placing_upper_limits}

\begin{figure}
    \centering
    \includegraphics[width=\linewidth]{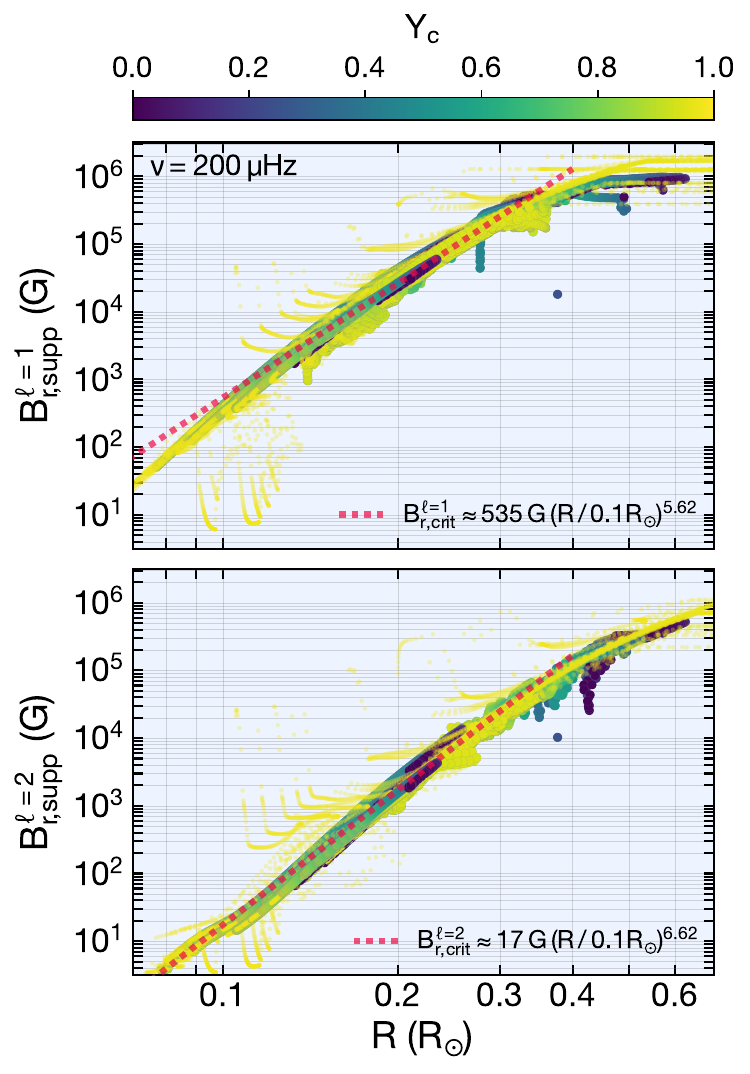}
    \caption{Minimum $B_{r,\mathrm{crit}}$ attained within the g-mode cavity for a $\nu=200\,\mu\mathrm{Hz}$ g mode with $\ell=1$ (\textit{top}) and $\ell=2$ (\textit{bottom}) for every sdB stellar model in our grid.
    Power law fits restricted to $R<0.4R_\odot$ and $Y_c\leq0.95$, and young sdB models with $Y_c>0.95$ are shown in with smaller, translucent scatter points for visual clarity.
    }
    \label{fig:FIG_sdb_size_dependence}
\end{figure}

In this Section, we summarize our method for deriving asteroseismic upper limits on the magnetic field using the non-detection of magnetic effects.
We closely follow the methodology of \citet{Rui:2025:wdmagnet} and \citet{Blatman:2025:ultramagnet}.

The detection of a g mode implies a magnetic field too weak to suppress it.
Using Equation \ref{eqn:bcrit}, the measurement of a g mode with angular degree $\ell$ and frequency $\nu$ constrains $B_r$ as a function of radius:
\begin{equation} \label{eqn:bsupp_profile}
    B_r(r) \lesssim \frac{4\pi^{5/2}}{\sqrt{\ell(\ell+1)}}\frac{\sqrt{\rho}\nu^2r}{N}\mathrm{,}
\end{equation}
within the g-mode cavity.
Although Equation \ref{eqn:bsupp_profile} constrains the entire magnetic field profile, in most cases we find in our models that the upper limit on the near-surface field is significantly stronger.
In this case, Equation \ref{eqn:bsupp_profile} can be collapsed into an upper limit on the near-surface field:
\begin{equation}
    B_{r,\mathrm{surf}} \lesssim B_{r,\mathrm{supp}}^{\ell\nu} = \frac{4\pi^{5/2}}{\sqrt{\ell(\ell+1)}}\left(\frac{\sqrt{\rho}r}{N}\right)_{\mathrm{min}\,\mathrm{g,}\ell\nu}\nu^2\mathrm{,}
\end{equation}
where $(\sqrt{\rho}r/N)_{\mathrm{min}\,\mathrm{g,}\ell\nu}$ is the minimum value of $\sqrt{\rho}r/N$ over the g-mode cavity.
The magnetic-suppression-based upper limit is then the minimum such constraint derived from any observed g mode in a given star, i.e.,
\begin{equation}
    B_{r,\mathrm{supp}} = \min\left\lbrace B_{r,\mathrm{supp}}^{\ell\nu}\right\rbrace_{\mathrm{modes}}\mathrm{,}
\end{equation}
which is always set by one of the lowest-frequency modes for one of the observed angular degrees $\ell$.

A gravity wave of frequency $\nu$ can propagate within shells satisfying $\nu<S_\ell/2\pi$ and $\nu<N/2\pi$.
In essentially all cases we explore, the outer boundary of the g-mode cavity is set by the Lamb frequency $S_\ell\propto\sqrt{\ell(\ell+1)T_{\mathrm{eff}}}/R$, because our models do not develop thick near-surface convective zones at which the Brunt--V\"ais\"al\"a frequency would turn over \citep[this is called the $S_\ell$-limited case by][]{Rui:2025:wdmagnet}.
This behavior can be seen in the second row of Figure \ref{fig:FIG_compare_models}, and is analogous to the behavior of high-frequency g modes in white dwarfs \citep{Rui:2025:wdmagnet}.
Because $B_{r,\mathrm{crit}}$ decreases sharply near the surface, the minimum $B_{r,\mathrm{crit}}$ is usually attained at the outer boundary of the g-mode cavity.
Since $S_\ell$ is typically larger for more compact stars, this minimum value of $B_{r,\mathrm{crit}}$ is a strong increasing function of $R$.
This explains various trends we observe in our calculations, namely:
\begin{itemize}
    \item More massive sdBs tend to have higher minimum $B_{r,\mathrm{crit}}$ because they tend to have larger radii.
    \item sdBs with high hydrogen masses sufficient to significantly increase their radii tend to have higher minimum $B_{r,\mathrm{crit}}$.
    \item As sdBs evolve, their minimum $B_{r,\mathrm{crit}}$ tends to decrease as the radii gradually increases.
    In many cases, the minimum $B_{r,\mathrm{crit}}$ increases again at the end of the sdB's life due to a contraction in radius.
\end{itemize}
To illustrate this point, Figure \ref{fig:FIG_sdb_size_dependence} shows the minimum $B_{r,\mathrm{crit}}$ for both $\ell=1$ and $\ell=2$ for a g mode with typical sdB frequency $\nu=200\,\mu\mathrm{Hz}$ for every model on our grid, excluding very young sdB models with $Y_c>0.95$ (Section \ref{sec:models}).
Across all of these model profiles (which differ across total mass $M$, hydrogen mass $M_H$, and central helium fraction $Y_c$), the minimum $B_{r,\mathrm{crit}}$ behaves as a steep, relatively tight power law in $R$.
Exceptions to this trend occur for $R>0.4R_\odot$ as well as in models which form the downward scatter in this relation in Figure \ref{fig:FIG_sdb_size_dependence}.
In both of these cases, the minimum $B_{r,\mathrm{crit}}$ is instead achieved within the g-mode cavity, at the location of the sharply stratified hydrogen-to-helium transition layer.
Young sdB models are found to experience significant scatter around this trend, likely owing to significant helium enhancements in their outer layers which have not yet had time to settle.

\subsection{Bounding the magnetic fields of observed g-mode pulsating hot subdwarfs}
\label{sec:bounding_fields}

\begin{figure}
    \centering
    \includegraphics[width=\linewidth]{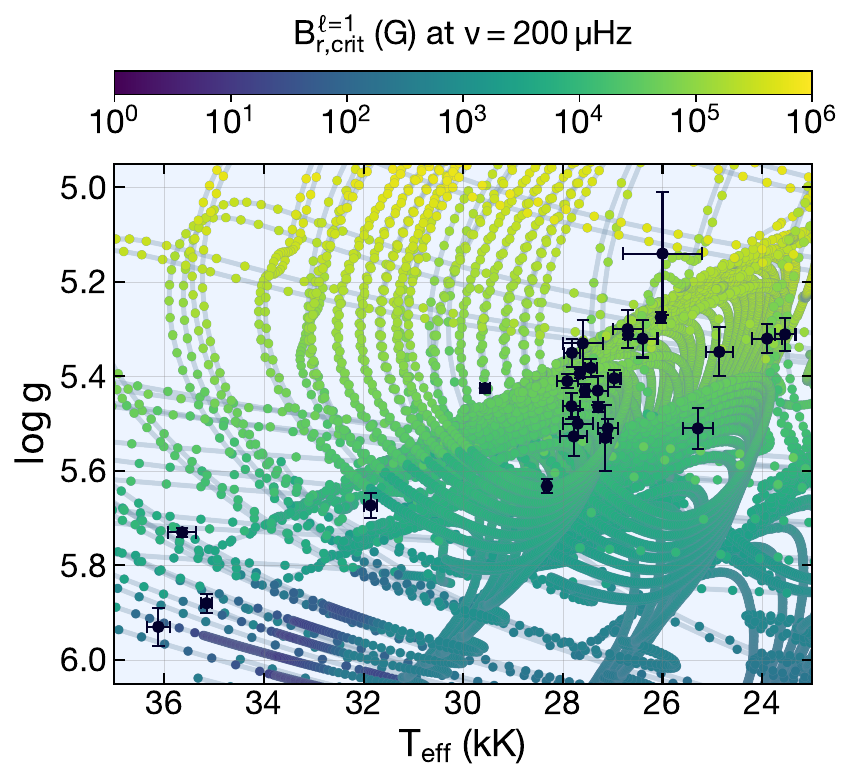}
    \caption{Kiel diagram showing observed g-mode pulsating sdBs considered in this work (black points) superposed over our stellar models, color-coded by the minimum value of $B_{r,\mathrm{crit}}$ attained within the g-mode cavity for a dipole ($\ell=1$) mode with $\nu=200\,\mu\mathrm{Hz}$.
    }
    \label{fig:FIG_observed_and_models}
\end{figure}

In this study, we seek to place conservative asteroseismic upper limits on magnetic fields in \NSDB{} observed g-mode pulsating hot subdwarfs, shown in Figure \ref{fig:FIG_observed_and_models} and summarized in Table \ref{tab:TAB_field_constraints}.
As low-frequency g modes are more sensitive to magnetism at a given $\ell$, we consider only the lowest-frequency $\ell=1$ and $\ell=2$ g modes for our proof-of-concept analysis.
To select these modes, we accept the mode identifications of the original observational references when presented as secure, although we acknowledge uncertainties within this process (which typically involves detecting rotational multiplets or g-mode period spacings).
In the case where a star has been observed by more than one study, we take the minimum $\ell=1$ and $\ell=2$ g-mode frequencies detected across all studies.
To be conservative, we assume that any modes with ambiguous identifications are dipole modes ($\ell=1$), since dipole modes are the least sensitive to magnetic fields.
Furthermore, we exclude modes reported as $\ell\geq3$; although in principle these are extraordinarily sensitive to magnetic fields, they are expected to be geometrically canceled and are observationally difficult to identify unambiguously.
We also ignore nonlinear combination frequencies, since they do not correspond to actual modes and are not expected to behave the same way as true g modes.
Finally, for simplicity, we ignore more complex pulsational behaviors, including mode trapping \citep{Uzundag:2017:sdBData,Cunha:2025:sdBClumpComparison}, amplitude/frequency variations \citep{Kern:2017:sdBData,Zong:2018:sdBData,Ma:2023:sdBData}, and nonlinear effects \citep[e.g.,][]{Zong:2016:sdBData}.

We assess the degree to which an observed sdB star matches a stellar model by comparing the star's observed spectroscopic parameters, $T_{\mathrm{eff}}\pm\sigma_{T_{\mathrm{eff}}}$ and $\log g\pm\sigma_{\log g}$, to those of the model, $T^{\mathrm{model}}_{\mathrm{eff}}$ and $(\log g)^{\mathrm{model}}$.
To do this, we compute
\begin{equation} \label{eqn:chisq}
    \chi^2 = \left(\frac{T_{\mathrm{eff}} - T^{\mathrm{model}}_{\mathrm{eff}}}{\sigma_{T_{\mathrm{eff}}}}\right)^2 + \left(\frac{\log g - (\log g)^{\mathrm{model}}}{\sigma_{\log g}}\right)^2\mathrm{,}
\end{equation}
and consider a stellar model a ``plausible'' fit to an observed star if $\chi^2<4$, which roughly corresponds to a $2\sigma$ tolerance on a disagreement between observed and model parameters.
For each plausibly fitting model, we compute $B^{\ell=1}_{r,\mathrm{supp}}$ and $B^{\ell=2}_{r,\mathrm{supp}}$ using the lowest-frequency $\ell=1$ and $\ell=2$ g modes (when available).
We conservatively place an upper limit on the magnetic field for $\ell=1$ and $\ell=2$ separately by taking the maximum value of $B^{\ell=1}_{r,\mathrm{supp}}$ and $B^{\ell=2}_{r,\mathrm{supp}}$ across all plausible models.
The final upper limit on the magnetic field is then given by $B_{r,\mathrm{supp}}=\min\left\lbrace B^{\ell=1}_{r,\mathrm{supp}},B^{\ell=2}_{r,\mathrm{supp}}\right\rbrace$.

We adapt our procedure slightly to accommodate the sdB star KIC 9472174 (also known as 2M1938+4603), which is in an eclipsing binary with an M dwarf star \citep{Oestensen:2010:sdBData}.
First, its tightly constrained spectroscopic parameters $T_{\mathrm{eff}}=(29564\pm106)\,\mathrm{kK}$ and $\log g=5.425\pm0.009$ reported by \citet{Oestensen:2010A:sdBData} are only consistent with a few model profiles with low masses $M\approx0.40M_\odot$ corresponding to a stellar progenitor with $M_{\mathrm{TRGB}}=3.0M_\odot$ on the tip of the red giant branch.
However, \citet{Oestensen:2010:sdBData} measure a more typical mass $M=(0.48\pm0.03)M_\odot$ using a combination of photometry and spectroscopy, including radial velocity information from the orbit.
To find reasonable models for this star, we therefore inflate the uncertainties by a factor of $3$ for the purposes of evaluating Equation \ref{eqn:chisq}, broadening the range of acceptable models to include more typical-mass sdBs with progenitor masses $M_{\mathrm{TRGB}}=1.2M_\odot$ and $M_{\mathrm{TRGB}}=3.5M_\odot$.
Second, \citet{Reed:2011:sdBData} identify an exceptionally low-frequency $\ell=2$ g mode with $\nu\approx50.28\,\mu\mathrm{Hz}$ which lies below any other quadrupole mode found in any other star in our sample, although this identification was made using an inconclusively determined period spacing.
This frequency lies below $S_2/2\pi$ for all of our model profiles that are consistent with the star's spectroscopic parameters, and, therefore, the outer boundary of the $\ell=2$ mode cavity is unphysically set by the outer boundary of the stellar model.
We therefore restrict our analysis for this star to the lowest-frequency $\ell=1$ mode only.

Although most of the \new{hot subdwarfs} in our sample are typical sdBs with H-rich atmospheres ($\log(n_{\mathrm{He}}/n_{\mathrm{H}})<-2$), three of them are intermediate-helium-rich \new{hot subdwarfs} ($\log(n_{\mathrm{He}}/n_{\mathrm{H}})>-1$) with strong enhancements in zirconium and other metals.
These apparently single \new{pulsating} subdwarfs, V366 Aqr \citep{Ahmad:2005:sdBData,Green:2011:sdBData,Randall:2015:sdBData}, Feige 46 \citep{Latour:2019:sdBData}, and PHL 417 \citep{Oestensen:2020:sdBData} \new{have unusually} high $T_{\mathrm{eff}}\gtrsim35\,\mathrm{kK}$\new{. This $T_{\mathrm{eff}}$ exceeds the temperature required to partially ionize the iron-group elements ($T\approx 20\,\mathrm{kK}$) that excite g mode pulsations through the conventional $\kappa$ mechanism.}
\new{Alternative proposed mechanisms for mode excitation include} mode instabilities due to the $\epsilon$ mechanism \citep{Battich:2018:EpsilonMechanism} \new{and a $\kappa$ mechanism induced by carbon/oxygen opacities in chemically enriched hot subdwarf envelopes \citep{Saio:2019:COkappamechanism}}.
Single \new{hot subdwarfs} with helium-rich atmospheres may be produced by single stars undergoing late core helium ignition due to strong winds \citep[``hot flashers'';][]{DCruz:1996:HotFlashers,Sweigart:1997:HotFlashers,MillerBertolami:2008:HotFlashers}, or in mergers between two helium white dwarfs \citep{Saio:2000:DoubleHeWDMergers,Zhang:2012:DoubleHeWDMergers,Schwab:2018:HeWDMerger} or between a helium white dwarf and a main-sequence star (\citealt{Clausen:2011:MSHeWDMergers,Zhang:2017:MSHeWDMergers}), although the outcome depends on the strength of red giant winds, see \citealt{Rui:2024:m_plus_wd}.
Intermediate helium-rich sdBs resulting from both single- and double-degenerate mergers can retain some hydrogen \citep{Hall:2016:HAtmospheresdBs,Zhang:2017:MSHeWDMergers}, which is expected to become hydrogen-rich a few megayear after the subdwarf is born due to gravitational settling \citep{Zhang:2017:MSHeWDMergers}.

All three stars coincide in the Kiel diagram with young sdB models with ages $\lesssim\mathrm{few}\,\mathrm{kyr}$ and order-unity surface helium fractions, similar to the observed intermediate helium-rich \new{hot subdwarf} stars.
However, because these objects are likely to have unusual formation histories \new{that are not captured in our evolutionary models}, our asteroseismic upper limits on the magnetic field should be considered especially uncertain for these stars. \new{Specifically, \citet{Saio:2019:COkappamechanism} and \citet{Oestensen:2020:sdBData} show that the excitation of g modes in these stars could be explained if their interior carbon/oxygen abundances are substantially enhanced.
If true, the peculiar structures of these stars would not be fully captured by our \texttt{MESA} model grids.}
Similar caveats apply to the other apparently single pulsating \new{hot subdwarfs} in our sample, which also ostensibly formed from helium white dwarf mergers.
Future work should investigate the asteroseismic sensitivity of more realistic merger remnant structures to near-surface magnetic fields.

\section{Results and discussion} \label{sec:results}

\begin{figure}
    \centering
    \includegraphics[width=0.48\textwidth]{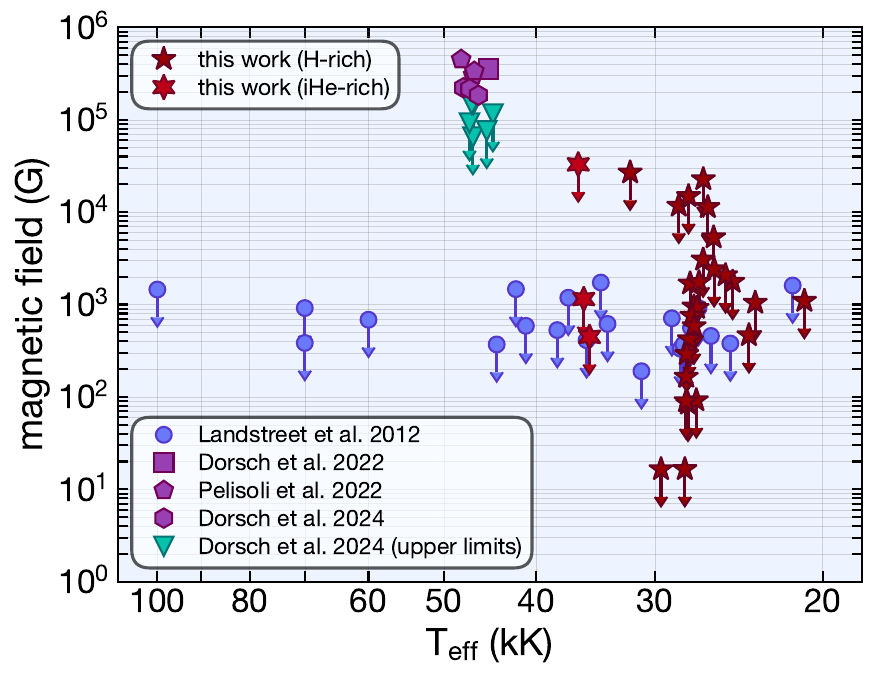}
    \caption{Asteroseismic upper limits on magnetism in \NSDB{} \new{hot subdwarf} stars using the non-detection of magnetic g-mode suppression.
    For reference, we also show the spectropolarimetric upper limits of \citet{Landstreet:2012:SpectropolSubdwarfs} on another sample of sdB and sdO stars, as well as the iHe-sdO stars of \citet{Dorsch:2022:MagneticsdO}, \citet{Pelisoli:2022:MagneticsdOs}, and \citet{Dorsch:2024:MagneticsdOs}, including the $7$ stars with confirmed magnetic fields.}
    \label{fig:FIG_field_constraints}
\end{figure}

Table \ref{tab:TAB_field_constraints} summarizes our asteroseismic upper limits on $B_r$ in our sample of \NSDB{} hot subdwarfs, as well as the parameters of the least-constraining (most conservative) stellar profiles used to estimate them.
These constraints span a very broad range $\sim20\,\mathrm{G}$--$30\,\mathrm{kG}$, which is a strong function of both the structure of the outer layers of the star and the lowest observed g-mode frequencies.
The strongest of these asteroseismic magnetic upper limits are similar to those placed by \citet{Rui:2025:wdmagnet} on pulsating white dwarfs.
This wide range of magnetic field constraints extends to the V366 Aquarid stars, for which we derive asteroseismic upper limits $B_r\lesssim400\,\mathrm{G}$, $30\,\mathrm{kG}$, and $1\,\mathrm{kG}$ in V366 Aqr, Feige 46, and PHL 417, respectively.
The least-constraining stellar profiles in Table \ref{tab:TAB_field_constraints} are biased towards lower $\log g$ and lower $T_{\mathrm{eff}}$ than the measured values, since we have conservatively selected the least constraining stellar models which tend to have larger radii.
However, asteroseismic upper limits can only be placed on g-mode pulsating hot subdwarfs, most of which have lower effective temperatures (all except 6 of the \new{hot subdwarfs} in our sample have $T_{\mathrm{eff}}<28\,\mathrm{kK}$).

Constraints on hot subdwarf magnetism are also available from spectroscopy and spectropolarimetry, although these methods probe different components of the field.
Spectropolarimetry is sensitive to weak values of the average longitudinal field $\langle B_z\rangle$, although it is only feasible for bright targets and is susceptible to geometric cancellation effects \citep[e.g.,][]{Bueno:2003:Spectropolarimetry}.
Figure \ref{fig:FIG_field_constraints} compares our asteroseismic constraints constraints to $2\sigma$ upper limits on $\langle B_z\rangle$ implied by the spectropolarimetric measurements of \citet{Landstreet:2012:SpectropolSubdwarfs}.
These spectropolarimetric upper limits much more consistently lie between a few hundred gauss and a few kilogauss.
Their sample of 24 hot subdwarfs (which is distinct from our own) includes both sdB and sdO stars and spans a wide range in effective temperatures $21.5\,\mathrm{kK}\leq T_{\mathrm{eff}}\leq100\,\mathrm{kK}$.
Spectroscopy, in turn, can measure comparatively high values of the average of the magnetic field modulus $\langle|\vec{B}|\rangle$ in a wider range of targets, although not as sensitively.
Figure \ref{fig:FIG_field_constraints} also shows the $7$ iHe-sdO stars with secure magnetic field detections based on the spectroscopic measurement of Zeeman-split hydrogen and helium lines \citep{Dorsch:2022:MagneticsdO,Dorsch:2024:MagneticsdOs,Pelisoli:2022:MagneticsdOs}.
These spectroscopic measurements place the absolute value of the magnetic field $\langle|\vec{B}|\rangle$ at hundreds of kilogauss.
We also include spectroscopic magnetic upper limits of \citet{Dorsch:2024:MagneticsdOs} on $6$ iHe-sdO stars which appear to share similar properties with the known magnetic stars, including a mysterious broad absorption feature at $\approx4630$\,\AA.
Because asteroseismology is sensitive to the radial component of the field $B_r$ (including its subsurface values), its combination with spectroscopy and spectropolarimetry may offer a more complete picture of the geometries of hot subdwarf magnetic fields.

Although differing in precisely which magnetic field properties they constrain, these three methods are consistent in affirming the absence of magnetism in ``normal'' sdBs.
In a handful of cases, our asteroseismic upper limits exclude the presence of even exceptionally weak fields $B_r\gtrsim10\mathrm{s}\,\mathrm{G}$, deepening this mystery.
Notably, our sample contains both apparently single \new{hot subdwarfs} and those with known stellar and white dwarf companions over a broad range of periods $0.1\,\mathrm{d}\lesssim P_{\mathrm{orb}}\lesssim30\,\mathrm{d}$ (as indicated in Table \ref{tab:TAB_field_constraints}), implying that \new{hot subdwarfs} fail to develop strong, long-lived magnetic fields under a variety of binary interactions.
All of the $7$ known magnetic hot subdwarfs belong to the rare iHe-sdO class, members of which are thought to form in double-degenerate mergers that are likely to produce strong magnetic fields.
However, as pointed out by previous authors \citep{Dorsch:2022:MagneticsdO,Dorsch:2024:MagneticsdOs,Pelisoli:2022:MagneticsdOs}, it is unclear why more typical He-sdO stars are not significantly magnetic when they are also thought to have formed in mergers.
In general, it remains unclear under what circumstances binary interactions are expected to leave long-lived magnetized remnants.
Merger-induced surface fields may either be very short-lived \citep{Dorsch:2022:MagneticsdO} or destroyed by turbulence in the merger remnant in all but those arising from certain merger configurations \citep{Pakmor:2024:HeWDMergers}.
Similarly, magnetic fields that are generated during common-envelope events \citet{Tout:2008:CEDynamo,Nordhaus:2011:CEEMagnetism} are expected to be initially confined to the outer layers and may dissipate soon after the envelope is removed \citep[cf.][]{Potter:2010:CEEMagnetismEvolution}.
The dearth of strong magnetic fields in hot subdwarfs may need to be reconciled with their observed rotation periods, for which magnetic braking may be a critical ingredient \citep{Moyano:2026:SubdwarfAccretion,Moyano2026}.

\subsection{Magnetic g-mode frequency shifts} \label{sec:freq_shifts}

\begin{figure*}
    \centering
    \includegraphics[width=\linewidth]{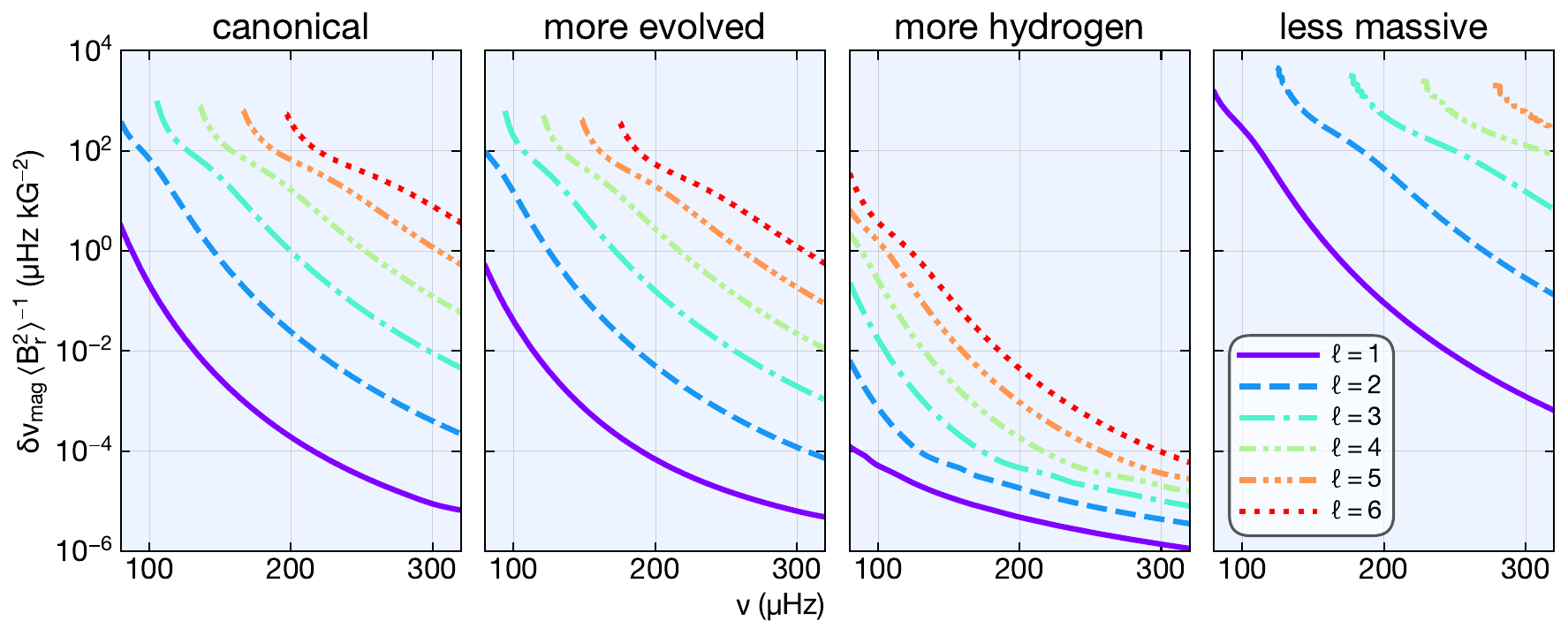}
    \caption{The magnitude of the average magnetic frequency shift $\delta\nu_{\mathrm{mag}}$ produced by a $1\,\mathrm{kG}$ magnetic field ($\langle B_r^2\rangle=1\,\mathrm{kG}^2$) for g modes of various angular degrees in sdB stellar models with different parameters.
    The stellar models used to generate these predictions are the same as those used in Figure \ref{fig:FIG_compare_models}.
    We omit frequencies at which the outer boundary of the g mode unphysically reaches all the way to the outermost cell of the stellar model.
    }
    \label{fig:FIG_freq_shifts}
\end{figure*}

Magnetic fields stronger than $B_{r,\mathrm{crit}}$ are expected to strongly suppress gravity-wave propagation.
However, weaker magnetic fields can still leave measurable impacts on g modes by modifying their frequencies.
Because the precise magnetic shift experienced by a g mode depends on $\ell$ and $m$, magnetic fields usually cause rotational multiplets to become ``asymmetric'' in the sense that modes in the multiplet become non-uniformly spaced in frequency \citep{Bugnet:2021:MagneticI,Li:2022:30to100kG}.

Although the magnetic shift of any individual mode depends on the magnetic field geometry, the average shift $\delta\nu_{\mathrm{mag}}$ experienced by modes within a multiplet with angular degree $\ell$ has a nice closed form from perturbation theory \citep{Bugnet:2021:MagneticI,Li:2022:30to100kG,Mathis:2023:MagneticAsymmetry,Das:2024:ComplexMagnetic}:
\begin{equation} \label{eqn:delta_nu_mag}
    \delta\nu_{\mathrm{mag}} = A_\ell\frac{\mathscr{I}}{64\pi^5\nu^3}\langle B_r^2\rangle\mathrm{,}
\end{equation}
where the magnetic sensitivity of the star \citep[e.g.,][]{Buchele:2026:MagneticSensitivity} is given by
\begin{equation} \label{eqn:curlyIasympt}
    \mathscr{I} \simeq \frac{\int_{\mathcal{R}^\ell_{\;\nu}}(N^3/\rho r^3)\,\mathrm{d}r}{\int_{\mathcal{R}^\ell_{\;\nu}}(N/r)\,\mathrm{d}r}\mathrm{.}
\end{equation}
The average $\langle B^2\rangle$ that appears in Equation \ref{eqn:delta_nu_mag} is given by
\begin{equation} \label{eqn:avgfield}
    \langle B_r^2\rangle = \int_{\mathcal{R}^\ell_{\;\nu}}\mathrm{d}r\,K(r)\int \frac{\mathrm{d}\Omega}{4\pi}\,B_r^2\mathrm{,}
\end{equation}
with a weighting kernel
\begin{equation} \label{eqn:weightfunction}
    K(r) \simeq \begin{cases}
        \frac{N^3/\rho r^3}{\int_{\mathcal{R}^\ell_{\;\nu}}(N^3/\rho r^3)\,\mathrm{d}r} & \mathrm{in}\,\mathcal{R}^\ell_{\;\nu} \\
        0 & \mathrm{otherwise}
    \end{cases}
\end{equation}
which is dominated by regions with low densities and strong stratification.

\citet{Rui:2025:wdmagnet} derive approximate asteroseismic magnetic upper limits in white dwarfs based on the non-detection of asymmetry in rotational triplets and quintuplets.
Although they find that asymmetry- and suppression-based asteroseismic upper limits are typically comparable, our preliminary tests indicate that the former are significantly less constraining in hot subdwarfs.
\citet{Das:2024:ComplexMagnetic} also find that rotational multiplets in magnetic stars can sometimes lack significant asymmetry, so we avoid reporting asymmetry-based upper limits for the stars in our sample.
Figure \ref{fig:FIG_freq_shifts} shows $\delta\nu_{\mathrm{mag}}$ as a function of frequency for four representative sdB stellar models.
Similarly to magnetic suppression effects, magnetic frequency shifts are larger for more compact stars, since both effects scale in analogous (though distinct) ways with $\rho$ and $N$ in the outer g-mode cavity.
In particular, especially strong sensitivity to magnetic fields is expected in high-$\ell$ modes, such as those reported in several hot subdwarfs \citep{Baran:2016:sdBData,Foster:2015:sdBData,Ma:2023:sdBData,Reed:2025:sdBData}.

The detection of magnetic frequency shifts would yield a direct and precise measurement of $\langle B_r^2\rangle$, rather than simply an upper limit.
If found, these magnetic frequency shifts would likely yield deeper insights into hot subdwarf magnetism, particularly if magnetic effects can be disentangled from the anomalous frequency variations which are sometimes observed \citep{Zong:2016:sdBData,Kern:2017:sdBData,Kern:2018:sdBData,Ma:2023:sdBData}.
Moreover, because magnetic frequency shifts encode the average of the magnetic field weighted by $K(r)$, they constrain the magnetic field profile in a complementary fashion to magnetic g-mode suppression.

\section{Conclusion}

In this work, we place asteroseismic upper limits on the magnetic fields of \NSDB{} pulsating \new{hot subdwarf} stars based on the \new{absence} of magnetic suppression \new{in} their lowest-frequency g modes.
These upper limits span a very broad range $\sim20\,\mathrm{G}$--$30\,\mathrm{kG}$ across systems, since the critical magnetic field required to suppress a g mode is a strong function of both its frequency and the star's properties.
In \new{hot subdwarfs} exhibiting g-mode pulsations, asteroseismic constraints on \new{hot subdwarf} magnetism are often competitive with or even significantly more sensitive than traditional methods based on spectroscopy or spectropolarimetry.
Our results reinforce those of previous studies reporting a lack of detectable magnetism in sdB stars and establish asteroseismology as a complementary method for probing hot subdwarf magnetism.

Our findings should be interpreted with several caveats.
First, the  Jeffreys-–Wentzel-–Kramers-–Brillouin (JWKB) approximation on which our estimates depend \citep[e.g.,][]{Fuller:2015:SuppressedDipole} assumes that the radial wavelength of the mode is much smaller than the length scale over which the background varies.
This assumption may break down in the outer layers of sdB stars in a way that is not yet theoretically understood.
Second, asteroseismology most strongly constrains the radial component of the magnetic field $B_r$ \citep[see, e.g.,][]{DuerfeldtPedros:2026:MagneticGDor}, and therefore cannot exclude strong, predominantly toroidal fields expected from many dynamo mechanisms that arise from differential rotation \citep[e.g.,][]{Spruit:2002:STDynamo,Fuller:2019:SlowingSpins}.
Finally, while there is strong observational evidence from pulsating red giants that magnetism significantly damps g modes \citep[see, e.g.,][]{Deheuvels:2023:MagneticRG,Deheuvels:2026:NearCritical}, some observational \citep{Mosser:2017:DepressedModes} and theoretical \citep{Loi:2020:MGPackets,Mueller:2025:RayTracing} studies find that some residual g-mode energy may survive.

Although Section \ref{sec:freq_shifts} presents rough estimates of the g-mode frequency shifts due to intermediate-strength magnetic fields, the topic of magnetic frequency shifts in hot subdwarfs merits deeper investigation.
If measured, magnetic frequency shifts may provide unique precision constraints on the strength and structure of \textit{internal} magnetic fields in sdB stars \citep[see, e.g., the theoretical works of][]{Mathis:2021:Magnetoasteroseismology,Bugnet:2021:MagneticI,Bugnet:2022:MagneticII,Mathis:2023:MagneticAsymmetry}.
Although magnetic g-mode frequency shifts have been reported in pulsating red giants \citep{Li:2022:30to100kG,Li:2023:13Magnetic,Deheuvels:2023:MagneticRG,Deheuvels:2026:NearCritical,Hatt:2024:MagneticRG,Villate:2026:MagneticOffset}, main-sequence stars \citep{Vandersnickt:2025:BcepMagnetic,Takata:2026:ToroidalGDor,Ihallaine:2026:MagneticGDor}, and a white dwarf \citep{Winget:1994:GD358}, none have been found in any pulsating hot subdwarf.
The detection and measurement of such frequency shifts may yield extraordinary insights into the conditions under which magnetic fields in hot subdwarfs form and evolve.

\section*{Acknowledgements}
\new{We thank the anonymous reviewer for their helpful feedback on this manuscript. }We thank Lucas Barrault, Daniel Blatman, Lisa Bugnet, Jim Fuller, Sivan Ginzburg, Ylva G\"otberg, and Eliot Quataert for useful comments and discussions, and Andy Baran for support regarding observations of one sdB star in our sample.
N.Z.R. acknowledges support from the NASA Hubble Fellowship grant HST-HF2-51589.001-A awarded by the Space Telescope Science Institute, which is operated by the Association of Universities for Research in Astronomy, Inc., for NASA, under contract NAS5-26555.
L.M. is supported by the Lyman Spitzer, Jr. Postdoctoral Fellowship of Princeton University, and the Gordon and Betty Moore Foundation Grant GBMF7392 at the Kavli Institute for Theoretical Physics (KITP). This work has also benefited from the Mitchel Postdoctoral Scholar Career Development Fund.

\section*{Data Availability}

\texttt{MESA} inlist files used to create our sdB stellar models can be found in a corresponding Zenodo upload \new{under an open-source 
Creative Commons Attribution license: 
\url{https://doi.org/10.5281/zenodo.23480966}}.

\bibliography{bibliography}{}

@ARTICLE{Paxton:2011:MESA,
  author = {{Paxton}, B. and {Bildsten}, L. and {Dotter}, A. and {Herwig}, F. and {Lesaffre}, P. and {Timmes}, F.},
  title = {{Modules for Experiments in Stellar Astrophysics (MESA)}},
  journal = {\apjs},
  archivePrefix = {arXiv},
  eprint = {1009.1622},
  primaryClass = {astro-ph.SR},
  year = {2011},
  month = {jan},
  volume = {192},
  eid = {3},
  pages = {3},
  doi = {10.1088/0067-0049/192/1/3},
  adsurl = {https://ui.adsabs.harvard.edu/abs/2011ApJS..192....3P},
}

@ARTICLE{Paxton:2013:MESA,
  author = {{Paxton}, B. and {Cantiello}, M. and {Arras}, P. and {Bildsten}, L. and {Brown}, E.~F. and {Dotter}, A. and {Mankovich}, C. and {Montgomery}, M.~H. and {Stello}, D. and {Timmes}, F.~X. and {Townsend}, R.},
  title = {{Modules for Experiments in Stellar Astrophysics (MESA): Planets, Oscillations, Rotation, and Massive Stars}},
  journal = {\apjs},
  archivePrefix = {arXiv},
  eprint = {1301.0319},
  primaryClass = {astro-ph.SR},
  year = {2013},
  month = {sep},
  volume = {208},
  eid = {4},
  pages = {4},
  doi = {10.1088/0067-0049/208/1/4},
  adsurl = {https://ui.adsabs.harvard.edu/abs/2013ApJS..208....4P},
}

@ARTICLE{Paxton:2015:MESA,
  author = {{Paxton}, B. and {Marchant}, P. and {Schwab}, J. and {Bauer}, E.~B. and {Bildsten}, L. and {Cantiello}, M. and {Dessart}, L. and {Farmer}, R. and {Hu}, H. and {Langer}, N. and {Townsend}, R.~H.~D. and {Townsley}, D.~M. and {Timmes}, F.~X.},
  title = {{Modules for Experiments in Stellar Astrophysics (MESA): Binaries, Pulsations, and Explosions}},
  journal = {\apjs},
  archivePrefix = {arXiv},
  eprint = {1506.03146},
  primaryClass = {astro-ph.SR},
  year = {2015},
  month = {sep},
  volume = {220},
  eid = {15},
  pages = {15},
  doi = {10.1088/0067-0049/220/1/15},
  adsurl = {https://ui.adsabs.harvard.edu/abs/2015ApJS..220...15P},
}

@ARTICLE{Paxton:2018:MESA,
  author = {{Paxton}, B. and {Schwab}, J. and {Bauer}, E.~B. and {Bildsten}, L. and {Blinnikov}, S. and {Duffell}, P. and {Farmer}, R. and {Goldberg}, J.~A. and {Marchant}, P. and {Sorokina}, E. and {Thoul}, A. and {Townsend}, R.~H.~D. and {Timmes}, F.~X.},
  title = {{Modules for Experiments in Stellar Astrophysics (MESA): Convective Boundaries, Element Diffusion, and Massive Star Explosions}},
  journal = {\apjs},
  archivePrefix = {arXiv},
  eprint = {1710.08424},
  primaryClass = {astro-ph.SR},
  year = {2018},
  month = {feb},
  volume = {234},
  eid = {34},
  pages = {34},
  doi = {10.3847/1538-4365/aaa5a8},
  adsurl = {https://ui.adsabs.harvard.edu/abs/2018ApJS..234...34P},
}

@ARTICLE{Paxton:2019:MESA,
       author = {{Paxton}, Bill and {Smolec}, R. and {Schwab}, Josiah and {Gautschy}, A. and
         {Bildsten}, Lars and {Cantiello}, Matteo and {Dotter}, Aaron and
         {Farmer}, R. and {Goldberg}, Jared A. and {Jermyn}, Adam S. and
         {Kanbur}, S.~M. and {Marchant}, Pablo and {Thoul}, Anne and
         {Townsend}, Richard H.~D. and {Wolf}, William M. and {Zhang}, Michael and
         {Timmes}, F.~X.},
        title = "{Modules for Experiments in Stellar Astrophysics (MESA): Pulsating Variable Stars, Rotation, Convective Boundaries, and Energy Conservation}",
      journal = {\apjs},
         year = "2019",
        month = "Jul",
       volume = {243},
       number = {1},
          eid = {10},
        pages = {10},
          doi = {10.3847/1538-4365/ab2241},
archivePrefix = {arXiv},
       eprint = {1903.01426},
 primaryClass = {astro-ph.SR},
       adsurl = {https://ui.adsabs.harvard.edu/abs/2019ApJS..243...10P}
}

@ARTICLE{Jermyn:2023:MESA,
       author = {{Jermyn}, Adam S. and {Bauer}, Evan B. and {Schwab}, Josiah and {Farmer}, R. and {Ball}, Warrick H. and {Bellinger}, Earl P. and {Dotter}, Aaron and {Joyce}, Meridith and {Marchant}, Pablo and {Mombarg}, Joey S.~G. and {Wolf}, William M. and {Sunny Wong}, Tin Long and {Cinquegrana}, Giulia C. and {Farrell}, Eoin and {Smolec}, R. and {Thoul}, Anne and {Cantiello}, Matteo and {Herwig}, Falk and {Toloza}, Odette and {Bildsten}, Lars and {Townsend}, Richard H.~D. and {Timmes}, F.~X.},
        title = "{Modules for Experiments in Stellar Astrophysics (MESA): Time-dependent Convection, Energy Conservation, Automatic Differentiation, and Infrastructure}",
      journal = {\apjs},
         year = 2023,
        month = mar,
       volume = {265},
       number = {1},
          eid = {15},
        pages = {15},
          doi = {10.3847/1538-4365/acae8d},
archivePrefix = {arXiv},
       eprint = {2208.03651},
 primaryClass = {astro-ph.SR},
       adsurl = {https://ui.adsabs.harvard.edu/abs/2023ApJS..265...15J}
}

@ARTICLE{Dorsch:2022:MagneticsdO,
       author = {{Dorsch}, M. and {Reindl}, N. and {Pelisoli}, I. and {Heber}, U. and {Geier}, S. and {Istrate}, A.~G. and {Justham}, S.},
        title = "{Discovery of a highly magnetic He-sdO star from a double-degenerate binary merger}",
      journal = {\aap},
         year = 2022,
        month = feb,
       volume = {658},
          eid = {L9},
        pages = {L9},
          doi = {10.1051/0004-6361/202142880},
archivePrefix = {arXiv},
       eprint = {2201.08146},
 primaryClass = {astro-ph.SR},
       adsurl = {https://ui.adsabs.harvard.edu/abs/2022A&A...658L...9D}
}

@ARTICLE{Pelisoli:2022:MagneticsdOs,
       author = {{Pelisoli}, Ingrid and {Dorsch}, M. and {Heber}, U. and {G{\"a}nsicke}, B. and {Geier}, S. and {Kupfer}, T. and {N{\'e}meth}, P. and {Scaringi}, S. and {Schaffenroth}, V.},
        title = "{Discovery and analysis of three magnetic hot subdwarf stars: evidence for merger-induced magnetic fields}",
      journal = {\mnras},
         year = 2022,
        month = sep,
       volume = {515},
       number = {2},
        pages = {2496-2510},
          doi = {10.1093/mnras/stac1069},
archivePrefix = {arXiv},
       eprint = {2204.06575},
 primaryClass = {astro-ph.SR},
       adsurl = {https://ui.adsabs.harvard.edu/abs/2022MNRAS.515.2496P}
}

@ARTICLE{Dorsch:2024:MagneticsdOs,
       author = {{Dorsch}, M. and {Jeffery}, C.~S. and {Philip Monai}, A. and {Tout}, C.~A. and {Snowdon}, E.~J. and {Monageng}, I. and {Scott}, L.~J.~A. and {Miszalski}, B. and {Woolf}, V.~M.},
        title = "{Discovery of three magnetic helium-rich hot subdwarfs with SALT}",
      journal = {\aap},
         year = 2024,
        month = nov,
       volume = {691},
          eid = {A165},
        pages = {A165},
          doi = {10.1051/0004-6361/202451306},
archivePrefix = {arXiv},
       eprint = {2410.02737},
 primaryClass = {astro-ph.SR},
       adsurl = {https://ui.adsabs.harvard.edu/abs/2024A&A...691A.165D}
}

@ARTICLE{Rui:2025:wdmagnet,
       author = {{Rui}, Nicholas Z. and {Fuller}, Jim and {Hermes}, J.~J.},
        title = "{Supersensitive Seismic Magnetometry of White Dwarfs}",
      journal = {\apj},
         year = 2025,
        month = mar,
       volume = {981},
       number = {1},
          eid = {72},
        pages = {72},
          doi = {10.3847/1538-4357/adaf9e},
archivePrefix = {arXiv},
       eprint = {2410.20557},
 primaryClass = {astro-ph.SR},
       adsurl = {https://ui.adsabs.harvard.edu/abs/2025ApJ...981...72R}
}

@ARTICLE{Fuller:2015:SuppressedDipole,
       author = {{Fuller}, Jim and {Cantiello}, Matteo and {Stello}, Dennis and {Garcia}, Rafael A. and {Bildsten}, Lars},
        title = "{Asteroseismology can reveal strong internal magnetic fields in red giant stars}",
      journal = {Science},
         year = 2015,
        month = oct,
       volume = {350},
       number = {6259},
        pages = {423-426},
          doi = {10.1126/science.aac6933},
archivePrefix = {arXiv},
       eprint = {1510.06960},
 primaryClass = {astro-ph.SR},
       adsurl = {https://ui.adsabs.harvard.edu/abs/2015Sci...350..423F}
}

@ARTICLE{Deheuvels:2023:MagneticRG,
       author = {{Deheuvels}, S. and {Li}, G. and {Ballot}, J. and {Ligni{\`e}res}, F.},
        title = "{Strong magnetic fields detected in the cores of 11 red giant stars using gravity-mode period spacings}",
      journal = {\aap},
         year = 2023,
        month = feb,
       volume = {670},
          eid = {L16},
        pages = {L16},
          doi = {10.1051/0004-6361/202245282},
archivePrefix = {arXiv},
       eprint = {2301.01308},
 primaryClass = {astro-ph.SR},
       adsurl = {https://ui.adsabs.harvard.edu/abs/2023A&A...670L..16D}
}

@article{Bagnulo:2021:MagneticWDs,
  title={New insight into the magnetism of degenerate stars from the analysis of a volume-limited sample of white dwarfs},
  author={Bagnulo, S and Landstreet, JD},
  journal={Monthly Notices of the Royal Astronomical Society},
  volume={507},
  number={4},
  pages={5902--5951},
  year={2021},
  publisher={Oxford University Press}
}

@ARTICLE{Vos:2021:PostMergerSubdwarf,
       author = {{Vos}, Joris and {Pelisoli}, Ingrid and {Budaj}, Jan and {Reindl}, Nicole and {Schaffenroth}, Veronika and {Bobrick}, Alexey and {Geier}, Stephan and {Hermes}, JJ and {Nemeth}, Peter and {{\O}stensen}, Roy and {Reding}, Joshua S. and {Uzundag}, Murat and {Vu{\v{c}}kovi{\'c}}, Maja},
        title = "{Looking into the cradle of the grave: J22564-5910, a potential young post-merger hot subdwarf}",
      journal = {\aap},
         year = 2021,
        month = nov,
       volume = {655},
          eid = {A43},
        pages = {A43},
          doi = {10.1051/0004-6361/202140391},
archivePrefix = {arXiv},
       eprint = {2106.03363},
 primaryClass = {astro-ph.SR},
       adsurl = {https://ui.adsabs.harvard.edu/abs/2021A&A...655A..43V}
}

@ARTICLE{Lignieres:2024:MagnetoGravitoinertial,
       author = {{Ligni{\`e}res}, F. and {Ballot}, J. and {Deheuvels}, S. and {Galoy}, M.},
        title = "{Perturbative analysis of the effect of a magnetic field on gravito-inertial modes}",
      journal = {\aap},
         year = 2024,
        month = mar,
       volume = {683},
          eid = {A2},
        pages = {A2},
          doi = {10.1051/0004-6361/202348243},
archivePrefix = {arXiv},
       eprint = {2311.13296},
 primaryClass = {astro-ph.SR},
       adsurl = {https://ui.adsabs.harvard.edu/abs/2024A&A...683A...2L}
}

@ARTICLE{Das:2024:ComplexMagnetic,
       author = {{Das}, S.~B. and {Einramhof}, L. and {Bugnet}, L.},
        title = "{Unveiling complex magnetic field configurations in red giant stars}",
      journal = {\aap},
         year = 2024,
        month = oct,
       volume = {690},
          eid = {A217},
        pages = {A217},
          doi = {10.1051/0004-6361/202450918},
archivePrefix = {arXiv},
       eprint = {2405.20133},
 primaryClass = {astro-ph.SR},
       adsurl = {https://ui.adsabs.harvard.edu/abs/2024A&A...690A.217D}
}

@ARTICLE{David:2025:Magnetogravity,
       author = {{David}, Cy S. and {Lecoanet}, Daniel and {Garaud}, Pascale},
        title = "{Conversion and Damping of Nonaxisymmetric Internal Gravity Waves in Magnetized Stellar Cores}",
      journal = {\apj},
         year = 2026,
        month = apr,
       volume = {1000},
       number = {2},
          eid = {292},
        pages = {292},
          doi = {10.3847/1538-4357/ae4d18},
archivePrefix = {arXiv},
       eprint = {2510.14026},
 primaryClass = {astro-ph.SR},
       adsurl = {https://ui.adsabs.harvard.edu/abs/2026ApJ..1000..292D}
}

@ARTICLE{Lecoanet:2017:MagneticConversion,
       author = {{Lecoanet}, D. and {Vasil}, G.~M. and {Fuller}, J. and {Cantiello}, M. and {Burns}, K.~J.},
        title = "{Conversion of internal gravity waves into magnetic waves}",
      journal = {\mnras},
         year = 2017,
        month = apr,
       volume = {466},
       number = {2},
        pages = {2181-2193},
          doi = {10.1093/mnras/stw3273},
archivePrefix = {arXiv},
       eprint = {1610.08506},
 primaryClass = {astro-ph.SR},
       adsurl = {https://ui.adsabs.harvard.edu/abs/2017MNRAS.466.2181L}
}

@ARTICLE{Rui:2024:TARM,
       author = {{Rui}, Nicholas Z. and {Ong}, J.~M. Joel and {Mathis}, St{\'e}phane},
        title = "{Asteroseismic g-mode period spacings in strongly magnetic rotating stars}",
      journal = {\mnras},
         year = 2024,
        month = jan,
       volume = {527},
       number = {3},
        pages = {6346-6362},
          doi = {10.1093/mnras/stad3461},
archivePrefix = {arXiv},
       eprint = {2310.19873},
 primaryClass = {astro-ph.SR},
       adsurl = {https://ui.adsabs.harvard.edu/abs/2024MNRAS.527.6346R}
}

@ARTICLE{Rui:2023:MagneticSuppression,
       author = {{Rui}, Nicholas Z. and {Fuller}, Jim},
        title = "{Gravity waves in strong magnetic fields}",
      journal = {\mnras},
         year = 2023,
        month = jul,
       volume = {523},
       number = {1},
        pages = {582-602},
          doi = {10.1093/mnras/stad1424},
archivePrefix = {arXiv},
       eprint = {2303.08147},
 primaryClass = {astro-ph.SR},
       adsurl = {https://ui.adsabs.harvard.edu/abs/2023MNRAS.523..582R}
}

@ARTICLE{Bugnet:2021:MagneticI,
       author = {{Bugnet}, L. and {Prat}, V. and {Mathis}, S. and {Astoul}, A. and {Augustson}, K. and {Garc{\'\i}a}, R.~A. and {Mathur}, S. and {Amard}, L. and {Neiner}, C.},
        title = "{Magnetic signatures on mixed-mode frequencies. I. An axisymmetric fossil field inside the core of red giants}",
      journal = {\aap},
         year = 2021,
        month = jun,
       volume = {650},
          eid = {A53},
        pages = {A53},
          doi = {10.1051/0004-6361/202039159},
archivePrefix = {arXiv},
       eprint = {2102.01216},
 primaryClass = {astro-ph.SR},
       adsurl = {https://ui.adsabs.harvard.edu/abs/2021A&A...650A..53B}
}

@ARTICLE{Vandersnickt:2025:BcepMagnetic,
       author = {{Vandersnickt}, Jelle and {Vanlaer}, Vincent and {Vanrespaille}, Mathijs and {Aerts}, Conny},
        title = "{Asteroseismic detection of an internal magnetic field in the B0.5V pulsator HD 192575}",
      journal = {\aap},
         year = 2025,
        month = dec,
       volume = {704},
          eid = {L13},
        pages = {L13},
          doi = {10.1051/0004-6361/202556850},
archivePrefix = {arXiv},
       eprint = {2511.21812},
 primaryClass = {astro-ph.SR},
       adsurl = {https://ui.adsabs.harvard.edu/abs/2025A&A...704L..13V}
}

@ARTICLE{Takata:2026:ToroidalGDor,
       author = {{Takata}, Masao and {Murphy}, Simon J. and {Kurtz}, Donald W. and {Saio}, Hideyuki and {Shibahashi}, Hiromoto},
        title = "{Asteroseismic detection of a predominantly toroidal magnetic field in the deep interior of the main-sequence F star KIC 9244992}",
      journal = {\mnras},
         year = 2026,
        month = jan,
       volume = {545},
       number = {3},
          eid = {staf2153},
        pages = {staf2153},
          doi = {10.1093/mnras/staf2153},
archivePrefix = {arXiv},
       eprint = {2512.00786},
 primaryClass = {astro-ph.SR},
       adsurl = {https://ui.adsabs.harvard.edu/abs/2026MNRAS.545f2153T}
}

@ARTICLE{Villate:2026:MagneticOffset,
       author = {{Villate}, M. and {Deheuvels}, S. and {Ballot}, J.},
        title = "{Seismic detection of core magnetic fields in red giants using the gravity offset}",
      journal = {\aap},
         year = 2026,
        month = mar,
       volume = {707},
          eid = {A366},
        pages = {A366},
          doi = {10.1051/0004-6361/202558608},
archivePrefix = {arXiv},
       eprint = {2602.14570},
 primaryClass = {astro-ph.SR},
       adsurl = {https://ui.adsabs.harvard.edu/abs/2026A&A...707A.366V}
}

@ARTICLE{Bugnet:2022:MagneticII,
       author = {{Bugnet}, L.},
        title = "{Magnetic signatures on mixed-mode frequencies. II. Period spacings as a probe of the internal magnetism of red giants}",
      journal = {\aap},
         year = 2022,
        month = nov,
       volume = {667},
          eid = {A68},
        pages = {A68},
          doi = {10.1051/0004-6361/202243167},
archivePrefix = {arXiv},
       eprint = {2208.14954},
 primaryClass = {astro-ph.SR},
       adsurl = {https://ui.adsabs.harvard.edu/abs/2022A&A...667A..68B}
}

@ARTICLE{Hatt:2024:MagneticRG,
       author = {{Hatt}, Emily J. and {Ong}, J.~M. Joel and {Nielsen}, Martin B. and {Chaplin}, William J. and {Davies}, Guy R. and {Deheuvels}, S{\'e}bastien and {Ballot}, J{\'e}r{\^o}me and {Li}, Gang and {Bugnet}, Lisa},
        title = "{Asteroseismic signatures of core magnetism and rotation in hundreds of low-luminosity red giants}",
      journal = {\mnras},
         year = 2024,
        month = oct,
       volume = {534},
       number = {2},
        pages = {1060-1076},
          doi = {10.1093/mnras/stae2053},
archivePrefix = {arXiv},
       eprint = {2409.01157},
 primaryClass = {astro-ph.SR},
       adsurl = {https://ui.adsabs.harvard.edu/abs/2024MNRAS.534.1060H}
}

@ARTICLE{Li:2022:30to100kG,
       author = {{Li}, Gang and {Deheuvels}, S{\'e}bastien and {Ballot}, J{\'e}r{\^o}me and {Ligni{\`e}res}, Fran{\c{c}}ois},
        title = "{Magnetic fields of 30 to 100 kG in the cores of red giant stars}",
      journal = {\nat},
         year = 2022,
        month = oct,
       volume = {610},
       number = {7930},
        pages = {43-46},
          doi = {10.1038/s41586-022-05176-0},
archivePrefix = {arXiv},
       eprint = {2208.09487},
 primaryClass = {astro-ph.SR},
       adsurl = {https://ui.adsabs.harvard.edu/abs/2022Natur.610...43L}
}

@ARTICLE{Li:2023:13Magnetic,
       author = {{Li}, Gang and {Deheuvels}, S{\'e}bastien and {Li}, Tanda and {Ballot}, J{\'e}r{\^o}me and {Ligni{\`e}res}, Fran{\c{c}}ois},
        title = "{Internal magnetic fields in 13 red giants detected by asteroseismology}",
      journal = {\aap},
         year = 2023,
        month = dec,
       volume = {680},
          eid = {A26},
        pages = {A26},
          doi = {10.1051/0004-6361/202347260},
archivePrefix = {arXiv},
       eprint = {2309.13756},
 primaryClass = {astro-ph.SR},
       adsurl = {https://ui.adsabs.harvard.edu/abs/2023A&A...680A..26L}
}

@ARTICLE{Stello:2016:MagneticPrevalence,
       author = {{Stello}, Dennis and {Cantiello}, Matteo and {Fuller}, Jim and {Huber}, Daniel and {Garc{\'\i}a}, Rafael A. and {Bedding}, Timothy R. and {Bildsten}, Lars and {Silva Aguirre}, Victor},
        title = "{A prevalence of dynamo-generated magnetic fields in the cores of intermediate-mass stars}",
      journal = {\nat},
         year = 2016,
        month = jan,
       volume = {529},
       number = {7586},
        pages = {364-367},
          doi = {10.1038/nature16171},
archivePrefix = {arXiv},
       eprint = {1601.00004},
 primaryClass = {astro-ph.SR},
       adsurl = {https://ui.adsabs.harvard.edu/abs/2016Natur.529..364S}
}

@ARTICLE{Donati:2009:Zeeman,
       author = {{Donati}, J.-F. and {Landstreet}, J.~D.},
        title = "{Magnetic Fields of Nondegenerate Stars}",
      journal = {\araa},
         year = 2009,
        month = sep,
       volume = {47},
       number = {1},
        pages = {333-370},
          doi = {10.1146/annurev-astro-082708-101833},
archivePrefix = {arXiv},
       eprint = {0904.1938},
 primaryClass = {astro-ph.SR},
       adsurl = {https://ui.adsabs.harvard.edu/abs/2009ARA&A..47..333D}
}

@ARTICLE{Nordhaus:2011:CEEMagnetism,
       author = {{Nordhaus}, J. and {Wellons}, S. and {Spiegel}, D.~S. and {Metzger}, B.~D. and {Blackman}, E.~G.},
        title = "{Formation of high-field magnetic white dwarfs from common envelopes}",
      journal = {Proceedings of the National Academy of Science},
         year = 2011,
        month = feb,
       volume = {108},
       number = {8},
        pages = {3135-3140},
          doi = {10.1073/pnas.1015005108},
archivePrefix = {arXiv},
       eprint = {1010.1529},
 primaryClass = {astro-ph.SR},
       adsurl = {https://ui.adsabs.harvard.edu/abs/2011PNAS..108.3135N}
}

@ARTICLE{Schneider:2016:MergerMagnetism,
       author = {{Schneider}, F.~R.~N. and {Podsiadlowski}, Ph. and {Langer}, N. and {Castro}, N. and {Fossati}, L.},
        title = "{Rejuvenation of stellar mergers and the origin of magnetic fields in massive stars}",
      journal = {\mnras},
         year = 2016,
        month = apr,
       volume = {457},
       number = {3},
        pages = {2355-2365},
          doi = {10.1093/mnras/stw148},
archivePrefix = {arXiv},
       eprint = {1601.05084},
 primaryClass = {astro-ph.SR},
       adsurl = {https://ui.adsabs.harvard.edu/abs/2016MNRAS.457.2355S}
}

@ARTICLE{Battich:2018:EpsilonMechanism,
       author = {{Battich}, Tiara and {Miller Bertolami}, Marcelo M. and {C{\'o}rsico}, Alejandro H. and {Althaus}, Leandro G.},
        title = "{Pulsational instabilities driven by the $\epsilon$ mechanism in hot pre-horizontal branch stars. I. The hot-flasher scenario}",
      journal = {\aap},
         year = 2018,
        month = jun,
       volume = {614},
          eid = {A136},
        pages = {A136},
          doi = {10.1051/0004-6361/201731463},
archivePrefix = {arXiv},
       eprint = {1801.07287},
 primaryClass = {astro-ph.SR},
       adsurl = {https://ui.adsabs.harvard.edu/abs/2018A&A...614A.136B}
}

@ARTICLE{Saio:2000:DoubleHeWDMergers,
       author = {{Saio}, Hideyuki and {Jeffery}, C. Simon},
        title = "{The evolution of a rapidly accreting helium white dwarf to become a low-luminosity helium star}",
      journal = {\mnras},
         year = 2000,
        month = apr,
       volume = {313},
       number = {4},
        pages = {671-677},
          doi = {10.1046/j.1365-8711.2000.03221.x},
       adsurl = {https://ui.adsabs.harvard.edu/abs/2000MNRAS.313..671S}
}

@ARTICLE{Schneider:2019:MergerMagnetism,
       author = {{Schneider}, Fabian R.~N. and {Ohlmann}, Sebastian T. and {Podsiadlowski}, Philipp and {R{\"o}pke}, Friedrich K. and {Balbus}, Steven A. and {Pakmor}, R{\"u}diger and {Springel}, Volker},
        title = "{Stellar mergers as the origin of magnetic massive stars}",
      journal = {\nat},
         year = 2019,
        month = oct,
       volume = {574},
       number = {7777},
        pages = {211-214},
          doi = {10.1038/s41586-019-1621-5},
archivePrefix = {arXiv},
       eprint = {1910.14058},
 primaryClass = {astro-ph.SR},
       adsurl = {https://ui.adsabs.harvard.edu/abs/2019Natur.574..211S}
}

@ARTICLE{Justham:2011:HerichSubdwarfFormation,
       author = {{Justham}, Stephen and {Podsiadlowski}, Philipp and {Han}, Zhanwen},
        title = "{On the formation of single and binary helium-rich subdwarf O stars}",
      journal = {\mnras},
         year = 2011,
        month = jan,
       volume = {410},
       number = {2},
        pages = {984-993},
          doi = {10.1111/j.1365-2966.2010.17497.x},
archivePrefix = {arXiv},
       eprint = {1008.1584},
 primaryClass = {astro-ph.SR},
       adsurl = {https://ui.adsabs.harvard.edu/abs/2011MNRAS.410..984J}
}

@ARTICLE{Jeffery:2026:SALTHotSubdwarfs,
       author = {{Jeffery}, C.~S. and {Dorsch}, M. and {Philip Monai}, A. and {Snowdon}, E.~J. and {Monageng}, I. and {Miszalski}, B.},
        title = "{The SALT survey of helium-rich hot subdwarfs: final sample and classification}",
      journal = {\mnras},
         year = 2026,
        month = feb,
       volume = {546},
       number = {1},
          eid = {staf2202},
        pages = {staf2202},
          doi = {10.1093/mnras/staf2202},
archivePrefix = {arXiv},
       eprint = {2512.10529},
 primaryClass = {astro-ph.SR},
       adsurl = {https://ui.adsabs.harvard.edu/abs/2026MNRAS.546f2202J}
}

@ARTICLE{Culpan:2022:HotSubdwarfCatalog,
       author = {{Culpan}, R. and {Geier}, S. and {Reindl}, N. and {Pelisoli}, I. and {Gentile Fusillo}, N. and {Vorontseva}, A.},
        title = "{The population of hot subdwarf stars studied with Gaia. IV. Catalogues of hot subluminous stars based on Gaia EDR3}",
      journal = {\aap},
         year = 2022,
        month = jun,
       volume = {662},
          eid = {A40},
        pages = {A40},
          doi = {10.1051/0004-6361/202243337},
archivePrefix = {arXiv},
       eprint = {2203.07938},
 primaryClass = {astro-ph.SR},
       adsurl = {https://ui.adsabs.harvard.edu/abs/2022A&A...662A..40C}
}

@ARTICLE{Schaffenroth:2022:sdBBinaries,
       author = {{Schaffenroth}, V. and {Pelisoli}, I. and {Barlow}, B.~N. and {Geier}, S. and {Kupfer}, T.},
        title = "{Hot subdwarfs in close binaries observed from space. I. Orbital, atmospheric, and absolute parameters, and the nature of their companions}",
      journal = {\aap},
         year = 2022,
        month = oct,
       volume = {666},
          eid = {A182},
        pages = {A182},
          doi = {10.1051/0004-6361/202244214},
archivePrefix = {arXiv},
       eprint = {2207.02001},
 primaryClass = {astro-ph.SR},
       adsurl = {https://ui.adsabs.harvard.edu/abs/2022A&A...666A.182S}
}

@ARTICLE{Xiong:2017:HotSubdwarfCEE,
       author = {{Xiong}, H. and {Chen}, X. and {Podsiadlowski}, Ph. and {Li}, Y. and {Han}, Z.},
        title = "{Subdwarf B stars from the common envelope ejection channel}",
      journal = {\aap},
         year = 2017,
        month = mar,
       volume = {599},
          eid = {A54},
        pages = {A54},
          doi = {10.1051/0004-6361/201629622},
archivePrefix = {arXiv},
       eprint = {1608.08739},
 primaryClass = {astro-ph.SR},
       adsurl = {https://ui.adsabs.harvard.edu/abs/2017A&A...599A..54X}
}

@ARTICLE{Zhang:2012:DoubleHeWDMergers,
       author = {{Zhang}, Xianfei and {Jeffery}, C. Simon},
        title = "{Evolutionary models for double helium white dwarf mergers and the formation of helium-rich hot subdwarfs}",
      journal = {\mnras},
         year = 2012,
        month = jan,
       volume = {419},
       number = {1},
        pages = {452-464},
          doi = {10.1111/j.1365-2966.2011.19711.x},
       adsurl = {https://ui.adsabs.harvard.edu/abs/2012MNRAS.419..452Z}
}

@ARTICLE{Clausen:2011:MSHeWDMergers,
       author = {{Clausen}, Drew and {Wade}, Richard A.},
        title = "{How to Make a Singleton sdB Star via Accelerated Stellar Evolution}",
      journal = {\apjl},
         year = 2011,
        month = jun,
       volume = {733},
       number = {2},
          eid = {L42},
        pages = {L42},
          doi = {10.1088/2041-8205/733/2/L42},
archivePrefix = {arXiv},
       eprint = {1104.5026},
 primaryClass = {astro-ph.SR},
       adsurl = {https://ui.adsabs.harvard.edu/abs/2011ApJ...733L..42C}
}

@ARTICLE{DCruz:1996:HotFlashers,
       author = {{D'Cruz}, Noella L. and {Dorman}, Ben and {Rood}, Robert T. and {O'Connell}, Robert W.},
        title = "{The Origin of Extreme Horizontal Branch Stars}",
      journal = {\apj},
         year = 1996,
        month = jul,
       volume = {466},
        pages = {359},
          doi = {10.1086/177515},
archivePrefix = {arXiv},
       eprint = {astro-ph/9511017},
 primaryClass = {astro-ph},
       adsurl = {https://ui.adsabs.harvard.edu/abs/1996ApJ...466..359D}
}

@ARTICLE{Sweigart:1997:HotFlashers,
       author = {{Sweigart}, Allen V.},
        title = "{Effects of Helium Mixing on the Evolution of Globular Cluster Stars}",
      journal = {\apjl},
         year = 1997,
        month = jan,
       volume = {474},
       number = {1},
        pages = {L23-L26},
          doi = {10.1086/310414},
       adsurl = {https://ui.adsabs.harvard.edu/abs/1997ApJ...474L..23S}
}

@INPROCEEDINGS{Bueno:2003:Spectropolarimetry,
       author = {{Trujillo Bueno}, J.},
        title = "{Astrophysical Spectropolarimetry and Magnetic Field Diagnostics}",
    booktitle = {Modelling of Stellar Atmospheres},
         year = 2003,
       editor = {{Piskunov}, N. and {Weiss}, W.~W. and {Gray}, D.~F.},
       series = {IAU Symposium},
       volume = {210},
        month = jan,
        pages = {243},
       adsurl = {https://ui.adsabs.harvard.edu/abs/2003IAUS..210..243T}
}

@ARTICLE{Scherbak:2023:CommonEnvelope,
  author = {{Scherbak}, Peter and {Fuller}, Jim},
  title = "{White dwarf binaries suggest a common envelope efficiency {\ensuremath{\alpha\sim}} 1/3}",
  journal = {\mnras},
  year = 2023,
  month = jan,
  volume = {518},
  number = {3},
  pages = {3966-3984},
  doi = {10.1093/mnras/stac3313},
archivePrefix = {arXiv},
  eprint = {2211.02036},
  primaryClass = {astro-ph.SR},
  adsurl = {https://ui.adsabs.harvard.edu/abs/2023MNRAS.518.3966S}
}

@ARTICLE{Kern:2017:sdBData,
       author = {{Kern}, J.~W. and {Reed}, M.~D. and {Baran}, A.~S. and {{\O}stensen}, R.~H. and {Telting}, J.~H.},
        title = "{Kepler observations of the pulsating subdwarf B star KIC 2697388: the detection of converging frequency multiplets in the full data set}",
      journal = {\mnras},
         year = 2017,
        month = feb,
       volume = {465},
       number = {1},
        pages = {1057-1065},
          doi = {10.1093/mnras/stw2794},
       adsurl = {https://ui.adsabs.harvard.edu/abs/2017MNRAS.465.1057K}
}

@ARTICLE{Vaisala:1925:BruntVaisala,
       author = {{V{\"a}is{\"a}l{\"a}}, Vilho},
        title = "{{\"U}ber die Wirkung der Windschwankungen auf die Pilotbeobachtungen}",
      journal = {Societus Scientiarum Fennica Commentationes Physico-Mathematicae},
         year = 1925,
        month = jan,
       volume = {2},
        pages = {19},
       adsurl = {https://ui.adsabs.harvard.edu/abs/1925SSFCP...2...19V}
}

@ARTICLE{Buchele:2026:MagneticSensitivity,
       author = {{Buchele}, L. and {Bugnet}, L. and {Muntean}, N. and {Hatt}, E.},
        title = "{Constraining the model-based uncertainties of asteroseismic magnetic field measurements in red giants}",
      journal = {\aap},
         year = 2026,
        month = sep,
       volume = {713},
          eid = {A294},
        pages = {A294},
          doi = {10.1051/0004-6361/202660897},
archivePrefix = {arXiv},
       eprint = {2607.00663},
 primaryClass = {astro-ph.SR},
       adsurl = {https://ui.adsabs.harvard.edu/abs/2026A&A...713A.294B}
}

@ARTICLE{Mathis:2023:MagneticAsymmetry,
       author = {{Mathis}, S. and {Bugnet}, L.},
        title = "{Asymmetries of frequency splittings of dipolar mixed modes: A window on the topology of deep magnetic fields}",
      journal = {\aap},
         year = 2023,
        month = aug,
       volume = {676},
          eid = {L9},
        pages = {L9},
          doi = {10.1051/0004-6361/202346832},
archivePrefix = {arXiv},
       eprint = {2306.11587},
 primaryClass = {astro-ph.SR},
       adsurl = {https://ui.adsabs.harvard.edu/abs/2023A&A...676L...9M}
}

@ARTICLE{Mathis:2021:Magnetoasteroseismology,
       author = {{Mathis}, S. and {Bugnet}, L. and {Prat}, V. and {Augustson}, K. and {Mathur}, S. and {Garcia}, R.~A.},
        title = "{Probing the internal magnetism of stars using asymptotic magneto-asteroseismology}",
      journal = {\aap},
         year = 2021,
        month = mar,
       volume = {647},
          eid = {A122},
        pages = {A122},
          doi = {10.1051/0004-6361/202039180},
archivePrefix = {arXiv},
       eprint = {2012.11050},
 primaryClass = {astro-ph.SR},
       adsurl = {https://ui.adsabs.harvard.edu/abs/2021A&A...647A.122M}
}

@ARTICLE{Ihallaine:2026:MagneticGDor,
       author = {{Ihallaine}, S. and {Ballot}, J. and {Ligni{\`e}res}, F. and {Ferri{\'e}}, L. and {Charpinet}, S. and {Galoy}, M. and {Li}, G.},
        title = "{Seismic signature of a magnetic field in the {\ensuremath{\gamma}} Doradus star KIC 2309579}",
      journal = {\aap},
         year = 2026,
        month = jun,
       volume = {710},
          eid = {L18},
        pages = {L18},
          doi = {10.1051/0004-6361/202660461},
archivePrefix = {arXiv},
       eprint = {2605.22533},
 primaryClass = {astro-ph.SR},
       adsurl = {https://ui.adsabs.harvard.edu/abs/2026A&A...710L..18I}
}

@ARTICLE{Winget:1994:GD358,
       author = {{Winget}, D.~E. and {Nather}, R.~E. and {Clemens}, J.~C. and {Provencal}, J.~L. and {Kleinman}, S.~J. and {Bradley}, P.~A. and {Claver}, C.~F. and {Dixson}, J.~S. and {Montgomery}, M.~H. and {Hansen}, C.~J. and {Hine}, B.~P. and {Birch}, P. and {Candy}, M. and {Marar}, T.~M.~K. and {Seetha}, S. and {Ashoka}, B.~N. and {Leibowitz}, E.~M. and {O'Donoghue}, D. and {Warner}, B. and {Buckley}, D.~A.~H. and {Tripe}, P. and {Vauclair}, G. and {Dolez}, N. and {Chevreton}, M. and {Serre}, T. and {Garrido}, R. and {Kepler}, S.~O. and {Kanaan}, A. and {Augusteijn}, T. and {Wood}, M.~A. and {Bergeron}, P. and {Grauer}, A.~D.},
        title = "{Whole Earth Telescope Observations of the DBV White Dwarf GD 358}",
      journal = {\apj},
         year = 1994,
        month = aug,
       volume = {430},
        pages = {839},
          doi = {10.1086/174455},
       adsurl = {https://ui.adsabs.harvard.edu/abs/1994ApJ...430..839W}
}

@ARTICLE{Deheuvels:2026:NearCritical,
       author = {{Deheuvels}, S. and {Ballot}, J. and {Ligni{\`e}res}, F. and {Li}, G. and {Villate}, M.},
        title = "{Near-critical magnetic fields in Kepler red giants}",
      journal = {\aap},
         year = 2026,
        month = may,
       volume = {709},
          eid = {A224},
        pages = {A224},
          doi = {10.1051/0004-6361/202659344},
archivePrefix = {arXiv},
       eprint = {2604.09901},
 primaryClass = {astro-ph.SR},
       adsurl = {https://ui.adsabs.harvard.edu/abs/2026A&A...709A.224D}
}

@ARTICLE{Mosser:2017:DepressedModes,
       author = {{Mosser}, B. and {Belkacem}, K. and {Pin{\c{c}}on}, C. and {Takata}, M. and {Vrard}, M. and {Barban}, C. and {Goupil}, M.-J. and {Kallinger}, T. and {Samadi}, R.},
        title = "{Dipole modes with depressed amplitudes in red giants are mixed modes}",
      journal = {\aap},
         year = 2017,
        month = feb,
       volume = {598},
          eid = {A62},
        pages = {A62},
          doi = {10.1051/0004-6361/201629494},
archivePrefix = {arXiv},
       eprint = {1610.03872},
 primaryClass = {astro-ph.SR},
       adsurl = {https://ui.adsabs.harvard.edu/abs/2017A&A...598A..62M}
}

@ARTICLE{Fuller:2019:SlowingSpins,
       author = {{Fuller}, Jim and {Piro}, Anthony L. and {Jermyn}, Adam S.},
        title = "{Slowing the spins of stellar cores}",
      journal = {\mnras},
         year = 2019,
        month = may,
       volume = {485},
       number = {3},
        pages = {3661-3680},
          doi = {10.1093/mnras/stz514},
archivePrefix = {arXiv},
       eprint = {1902.08227},
 primaryClass = {astro-ph.SR},
       adsurl = {https://ui.adsabs.harvard.edu/abs/2019MNRAS.485.3661F}
}

@ARTICLE{Spruit:2002:STDynamo,
       author = {{Spruit}, H.~C.},
        title = "{Dynamo action by differential rotation in a stably stratified stellar interior}",
      journal = {\aap},
         year = 2002,
        month = jan,
       volume = {381},
        pages = {923-932},
          doi = {10.1051/0004-6361:20011465},
archivePrefix = {arXiv},
       eprint = {astro-ph/0108207},
 primaryClass = {astro-ph},
       adsurl = {https://ui.adsabs.harvard.edu/abs/2002A&A...381..923S}
}

@ARTICLE{Loi:2020:MGPackets,
       author = {{Loi}, Shyeh Tjing},
        title = "{Magneto-gravity wave packet dynamics in strongly magnetized cores of evolved stars}",
      journal = {\mnras},
         year = 2020,
        month = apr,
       volume = {493},
       number = {4},
        pages = {5726-5742},
          doi = {10.1093/mnras/staa581},
archivePrefix = {arXiv},
       eprint = {2002.11130},
 primaryClass = {astro-ph.SR},
       adsurl = {https://ui.adsabs.harvard.edu/abs/2020MNRAS.493.5726L}
}

@ARTICLE{Mueller:2025:RayTracing,
       author = {{M{\"u}ller}, Jonas and {Copp{\'e}e}, Quentin and {Hekker}, Saskia},
        title = "{Oscillations of red giant stars with magnetic damping in the core: I. Dissipation of mode energy in dipole-like magnetic fields}",
      journal = {\aap},
         year = 2025,
        month = apr,
       volume = {696},
          eid = {A134},
        pages = {A134},
          doi = {10.1051/0004-6361/202553888},
archivePrefix = {arXiv},
       eprint = {2503.11451},
 primaryClass = {astro-ph.SR},
       adsurl = {https://ui.adsabs.harvard.edu/abs/2025A&A...696A.134M}
}

@ARTICLE{Heber:2009:sdBReview,
       author = {{Heber}, Ulrich},
        title = "{Hot Subdwarf Stars}",
      journal = {\araa},
         year = 2009,
        month = sep,
       volume = {47},
       number = {1},
        pages = {211-251},
          doi = {10.1146/annurev-astro-082708-101836},
       adsurl = {https://ui.adsabs.harvard.edu/abs/2009ARA&A..47..211H}
}

@ARTICLE{MillerBertolami:2008:HotFlashers,
       author = {{Miller Bertolami}, M.~M. and {Althaus}, L.~G. and {Unglaub}, K. and {Weiss}, A.},
        title = "{Modeling He-rich subdwarfs through the hot-flasher scenario}",
      journal = {\aap},
         year = 2008,
        month = nov,
       volume = {491},
       number = {1},
        pages = {253-265},
          doi = {10.1051/0004-6361:200810373},
archivePrefix = {arXiv},
       eprint = {0808.3580},
 primaryClass = {astro-ph},
       adsurl = {https://ui.adsabs.harvard.edu/abs/2008A&A...491..253M}
}

@ARTICLE{Schwab:2018:HeWDMerger,
       author = {{Schwab}, Josiah},
        title = "{Hot subdwarfs formed from the merger of two He white dwarfs}",
      journal = {\mnras},
         year = 2018,
        month = jun,
       volume = {476},
       number = {4},
        pages = {5303-5311},
          doi = {10.1093/mnras/sty586},
archivePrefix = {arXiv},
       eprint = {1803.00576},
 primaryClass = {astro-ph.SR},
       adsurl = {https://ui.adsabs.harvard.edu/abs/2018MNRAS.476.5303S}
}

@ARTICLE{Beloborodov:2014:MagneticOutbursts,
       author = {{Beloborodov}, Andrei M.},
        title = "{Magnetically powered outbursts from white dwarf mergers}",
      journal = {\mnras},
         year = 2014,
        month = feb,
       volume = {438},
       number = {1},
        pages = {169-176},
          doi = {10.1093/mnras/stt2140},
archivePrefix = {arXiv},
       eprint = {1311.0668},
 primaryClass = {astro-ph.HE},
       adsurl = {https://ui.adsabs.harvard.edu/abs/2014MNRAS.438..169B}
}

@ARTICLE{Tout:2008:CEDynamo,
       author = {{Tout}, C.~A. and {Wickramasinghe}, D.~T. and {Liebert}, J. and {Ferrario}, L. and {Pringle}, J.~E.},
        title = "{Binary star origin of high field magnetic white dwarfs}",
      journal = {\mnras},
         year = 2008,
        month = jun,
       volume = {387},
       number = {2},
        pages = {897-901},
          doi = {10.1111/j.1365-2966.2008.13291.x},
archivePrefix = {arXiv},
       eprint = {0805.0115},
 primaryClass = {astro-ph},
       adsurl = {https://ui.adsabs.harvard.edu/abs/2008MNRAS.387..897T}
}

@ARTICLE{Han:2002:sdBFormation,
       author = {{Han}, Z. and {Podsiadlowski}, Ph. and {Maxted}, P.~F.~L. and {Marsh}, T.~R. and {Ivanova}, N.},
        title = "{The origin of subdwarf B stars - I. The formation channels}",
      journal = {\mnras},
         year = 2002,
        month = oct,
       volume = {336},
       number = {2},
        pages = {449-466},
          doi = {10.1046/j.1365-8711.2002.05752.x},
archivePrefix = {arXiv},
       eprint = {astro-ph/0206130},
 primaryClass = {astro-ph},
       adsurl = {https://ui.adsabs.harvard.edu/abs/2002MNRAS.336..449H}
}

@ARTICLE{Vos:2020:sdBSMT,
       author = {{Vos}, J. and {Bobrick}, A. and {Vu{\v{c}}kovi{\'c}}, M.},
        title = "{Observed binary populations reflect the Galactic history. Explaining the orbital period-mass ratio relation in wide hot subdwarf binaries}",
      journal = {\aap},
         year = 2020,
        month = sep,
       volume = {641},
          eid = {A163},
        pages = {A163},
          doi = {10.1051/0004-6361/201937195},
archivePrefix = {arXiv},
       eprint = {2003.05665},
 primaryClass = {astro-ph.SR},
       adsurl = {https://ui.adsabs.harvard.edu/abs/2020A&A...641A.163V}
}

@ARTICLE{Han:2003:sdBFormation,
       author = {{Han}, Z. and {Podsiadlowski}, Ph. and {Maxted}, P.~F.~L. and {Marsh}, T.~R.},
        title = "{The origin of subdwarf B stars - II}",
      journal = {\mnras},
         year = 2003,
        month = may,
       volume = {341},
       number = {2},
        pages = {669-691},
          doi = {10.1046/j.1365-8711.2003.06451.x},
archivePrefix = {arXiv},
       eprint = {astro-ph/0301380},
 primaryClass = {astro-ph},
       adsurl = {https://ui.adsabs.harvard.edu/abs/2003MNRAS.341..669H}
}

@ARTICLE{Maxted:2001:sdBBinarity,
       author = {{Maxted}, P.~F.~L. and {Heber}, U. and {Marsh}, T.~R. and {North}, R.~C.},
        title = "{The binary fraction of extreme horizontal branch stars}",
      journal = {\mnras},
         year = 2001,
        month = oct,
       volume = {326},
       number = {4},
        pages = {1391-1402},
          doi = {10.1111/j.1365-2966.2001.04714.x},
archivePrefix = {arXiv},
       eprint = {astro-ph/0103342},
 primaryClass = {astro-ph},
       adsurl = {https://ui.adsabs.harvard.edu/abs/2001MNRAS.326.1391M}
}

@ARTICLE{Lecoanet:2022:MagneticSPB,
       author = {{Lecoanet}, Daniel and {Bowman}, Dominic M. and {Van Reeth}, Timothy},
        title = "{Asteroseismic inference of the near-core magnetic field strength in the main-sequence B star HD 43317}",
      journal = {\mnras},
         year = 2022,
        month = may,
       volume = {512},
       number = {1},
        pages = {L16-L20},
          doi = {10.1093/mnrasl/slac013},
archivePrefix = {arXiv},
       eprint = {2202.03440},
 primaryClass = {astro-ph.SR},
       adsurl = {https://ui.adsabs.harvard.edu/abs/2022MNRAS.512L..16L}
}

@ARTICLE{Blatman:2025:ultramagnet,
       author = {{Blatman}, Daniel and {Rui}, Nicholas Z. and {Ginzburg}, Sivan and {Fuller}, Jim},
        title = "{Seismology and diffusion of ultramassive white dwarf magnetic fields}",
      journal = {\mnras},
         year = 2025,
        month = sep,
       volume = {542},
       number = {3},
        pages = {2345-2352},
          doi = {10.1093/mnras/staf1365},
archivePrefix = {arXiv},
       eprint = {2507.05343},
 primaryClass = {astro-ph.SR},
       adsurl = {https://ui.adsabs.harvard.edu/abs/2025MNRAS.542.2345B}
}

@ARTICLE{Pakmor:2024:HeWDMergers,
       author = {{Pakmor}, R{\"u}diger and {Pelisoli}, Ingrid and {Justham}, Stephen and {Rajamuthukumar}, Abinaya S. and {R{\"o}pke}, Friedrich K. and {Schneider}, Fabian R.~N. and {de Mink}, Selma E. and {Ohlmann}, Sebastian T. and {Podsiadlowski}, Philipp and {Mor{\'a}n-Fraile}, Javier and {Vetter}, Marco and {Andrassy}, Robert},
        title = "{Large-scale ordered magnetic fields generated in mergers of helium white dwarfs}",
      journal = {\aap},
         year = 2024,
        month = nov,
       volume = {691},
          eid = {A179},
        pages = {A179},
          doi = {10.1051/0004-6361/202451352},
archivePrefix = {arXiv},
       eprint = {2407.02566},
 primaryClass = {astro-ph.SR},
       adsurl = {https://ui.adsabs.harvard.edu/abs/2024A&A...691A.179P}
}

@ARTICLE{Kilkenny:2007:PulsatingHotSubdwarfs,
       author = {{Kilkenny}, D.},
        title = "{Pulsating Hot Subdwarfs -- An Observational Review}",
      journal = {Communications in Asteroseismology},
         year = 2007,
        month = jun,
       volume = {150},
        pages = {234},
          doi = {10.1553/cia150s234},
       adsurl = {https://ui.adsabs.harvard.edu/abs/2007CoAst.150..234K}
}

@ARTICLE{Charpinet:1996:PulsatingsdBs,
       author = {{Charpinet}, S. and {Fontaine}, G. and {Brassard}, P. and {Dorman}, B.},
        title = "{The Potential of Asteroseismology for Hot, Subdwarf B Stars: A New Class of Pulsating Stars?}",
      journal = {\apjl},
         year = 1996,
        month = nov,
       volume = {471},
        pages = {L103},
          doi = {10.1086/310335},
archivePrefix = {arXiv},
       eprint = {astro-ph/9606069},
 primaryClass = {astro-ph},
       adsurl = {https://ui.adsabs.harvard.edu/abs/1996ApJ...471L.103C}
}

@ARTICLE{Ma:2024:sdBTidalSpinup,
       author = {{Ma}, Linhao and {Fuller}, Jim},
        title = "{Tidal Spin-up of Subdwarf B Stars}",
      journal = {\apj},
         year = 2024,
        month = nov,
       volume = {975},
       number = {1},
          eid = {1},
        pages = {1},
          doi = {10.3847/1538-4357/ad7788},
archivePrefix = {arXiv},
       eprint = {2408.16158},
 primaryClass = {astro-ph.SR},
       adsurl = {https://ui.adsabs.harvard.edu/abs/2024ApJ...975....1M}
}

@ARTICLE{Uzundag:2024:PulsatingHotSubdwarfs,
       author = {{Uzundag}, Murat and {Krzesinski}, Jurek and {Pelisoli}, Ingrid and {N{\'e}meth}, P{\'e}ter and {Silvotti}, Roberto and {Vu{\v{c}}kovi{\'c}}, Maja and {Dawson}, Harry and {Geier}, Stephan},
        title = "{A comprehensive search for hot subdwarf stars using Gaia and TESS. I. Pulsating hot subdwarf B stars}",
      journal = {\aap},
         year = 2024,
        month = apr,
       volume = {684},
          eid = {A118},
        pages = {A118},
          doi = {10.1051/0004-6361/202348829},
archivePrefix = {arXiv},
       eprint = {2401.17707},
 primaryClass = {astro-ph.SR},
       adsurl = {https://ui.adsabs.harvard.edu/abs/2024A&A...684A.118U}
}

@ARTICLE{Cantiello:2016:EvolvingMagnetic,
       author = {{Cantiello}, Matteo and {Fuller}, Jim and {Bildsten}, Lars},
        title = "{Asteroseismic Signatures of Evolving Internal Stellar Magnetic Fields}",
      journal = {\apj},
         year = 2016,
        month = jun,
       volume = {824},
       number = {1},
          eid = {14},
        pages = {14},
          doi = {10.3847/0004-637X/824/1/14},
archivePrefix = {arXiv},
       eprint = {1602.03056},
 primaryClass = {astro-ph.SR},
       adsurl = {https://ui.adsabs.harvard.edu/abs/2016ApJ...824...14C}
}

@ARTICLE{DuerfeldtPedros:2026:MagneticGDor,
       author = {{D{\"u}rfeldt-Pedros}, O. and {Antoci}, V. and {Lecoanet}, D. and {Guo}, Z. and {Labadie-Bartz}, J.},
        title = "{Near-core magnetic field strengths inferred from gravity modes in intermediate-mass stars}",
      journal = {\aap},
         year = 2026,
        month = jul,
       volume = {711},
          eid = {A197},
        pages = {A197},
          doi = {10.1051/0004-6361/202558765},
archivePrefix = {arXiv},
       eprint = {2606.12148},
 primaryClass = {astro-ph.SR},
       adsurl = {https://ui.adsabs.harvard.edu/abs/2026A&A...711A.197D}
}

@ARTICLE{Moyano2026,
       author = {{Moyano}, Facundo D. and {Liu}, Zhengwei and {Han}, Zhanwen and {Chen}, Xuefei and {Battich}, Tiara and {Podsiadlowski}, Philipp},
        title = "{Post-merger rotational evolution of magnetised single hot subdwarfs results in slow rotation due to magnetic braking}",
      journal = {arXiv e-prints},
         year = 2026,
        month = oct,
          eid = {arXiv:2610.19671},
        pages = {arXiv:2610.19671},
          doi = {10.48550/arXiv.2610.19671},
archivePrefix = {arXiv},
       eprint = {2610.19671},
 primaryClass = {astro-ph.SR},
       adsurl = {https://ui.adsabs.harvard.edu/abs/2026arXiv261019671M}
}

@ARTICLE{Moyano:2026:SubdwarfAccretion,
       author = {{Moyano}, Facundo D. and {Ge}, Hongwei and {Han}, Zhanwen and {Bordad{\'a}gua}, Beatriz and {Uzundag}, Murat and {Podsiadlowski}, Philipp and {Schaffenroth}, Veronika and {Chen}, Xuefei and {Liu}, Zhengwei},
        title = "{Asteroseismic rotation rates of hot subdwarf B stars hint at transient accretion from leftover common envelope matter}",
      journal = {\mnras},
         year = 2026,
        month = jun,
       volume = {548},
       number = {4},
          eid = {stag790},
        pages = {stag790},
          doi = {10.1093/mnras/stag790},
archivePrefix = {arXiv},
       eprint = {2604.22233},
 primaryClass = {astro-ph.SR},
       adsurl = {https://ui.adsabs.harvard.edu/abs/2026MNRAS.548ag790M}
}

@ARTICLE{Saio:2019:COkappamechanism,
       author = {{Saio}, H. and {Jeffery}, C.~S.},
        title = "{The excitation of g-mode pulsations in hot helium-rich subdwarfs}",
      journal = {\mnras},
         year = 2019,
        month = jan,
       volume = {482},
       number = {1},
        pages = {758-761},
          doi = {10.1093/mnras/sty2763},
       adsurl = {https://ui.adsabs.harvard.edu/abs/2019MNRAS.482..758S}
}

@ARTICLE{Loi:2017:AlfvenResonances,
       author = {{Loi}, Shyeh Tjing and {Papaloizou}, John C.~B.},
        title = "{Torsional Alfv{\'e}n resonances as an efficient damping mechanism for non-radial oscillations in red giant stars}",
      journal = {\mnras},
         year = 2017,
        month = may,
       volume = {467},
       number = {3},
        pages = {3212-3225},
          doi = {10.1093/mnras/stx281},
archivePrefix = {arXiv},
       eprint = {1701.08771},
 primaryClass = {astro-ph.SR},
       adsurl = {https://ui.adsabs.harvard.edu/abs/2017MNRAS.467.3212L}
}

@ARTICLE{Landstreet:2012:SpectropolSubdwarfs,
       author = {{Landstreet}, J.~D. and {Bagnulo}, S. and {Fossati}, L. and {Jordan}, S. and {O'Toole}, S.~J.},
        title = "{The magnetic fields of hot subdwarf stars}",
      journal = {\aap},
         year = 2012,
        month = may,
       volume = {541},
          eid = {A100},
        pages = {A100},
          doi = {10.1051/0004-6361/201219178},
archivePrefix = {arXiv},
       eprint = {1203.6815},
 primaryClass = {astro-ph.SR},
       adsurl = {https://ui.adsabs.harvard.edu/abs/2012A&A...541A.100L}
}

@ARTICLE{Pablo:2012:sdBData,
       author = {{Pablo}, Herbert and {Kawaler}, Steven D. and {Reed}, M.~D. and {Bloemen}, S. and {Charpinet}, S. and {Hu}, H. and {Telting}, J. and {{\O}stensen}, R.~H. and {Baran}, A.~S. and {Green}, E.~M. and {Hermes}, J.~J. and {Barclay}, T. and {O'Toole}, S.~J. and {Mullally}, Fergal and {Kurtz}, D.~W. and {Christensen-Dalsgaard}, J. and {Caldwell}, Douglas A. and {Christiansen}, Jessie L. and {Kinemuchi}, K.},
        title = "{Seismic evidence for non-synchronization in two close sdb+dM binaries from Kepler photometry}",
      journal = {\mnras},
         year = 2012,
        month = may,
       volume = {422},
       number = {2},
        pages = {1343-1351},
          doi = {10.1111/j.1365-2966.2012.20707.x},
archivePrefix = {arXiv},
       eprint = {1202.3649},
 primaryClass = {astro-ph.SR},
       adsurl = {https://ui.adsabs.harvard.edu/abs/2012MNRAS.422.1343P}
}

@ARTICLE{Telting:2014:sdBData,
       author = {{Telting}, J.~H. and {Baran}, A.~S. and {Nemeth}, P. and {{\O}stensen}, R.~H. and {Kupfer}, T. and {Macfarlane}, S. and {Heber}, U. and {Aerts}, C. and {Geier}, S.},
        title = "{KIC 7668647: a 14 day beaming sdB+WD binary with a pulsating subdwarf}",
      journal = {\aap},
         year = 2014,
        month = oct,
       volume = {570},
          eid = {A129},
        pages = {A129},
          doi = {10.1051/0004-6361/201424169},
archivePrefix = {arXiv},
       eprint = {1409.2345},
 primaryClass = {astro-ph.SR},
       adsurl = {https://ui.adsabs.harvard.edu/abs/2014A&A...570A.129T}
}

@ARTICLE{Baran:2016:sdBData,
       author = {{Baran}, A.~S. and {Telting}, J.~H. and {N{\'e}meth}, P. and {{\O}stensen}, R.~H. and {Reed}, M.~D. and {Kiaeerad}, F.},
        title = "{A subsynchronously rotating pulsating subdwarf B star in a short-period binary with a white dwarf companion}",
      journal = {\aap},
         year = 2016,
        month = jan,
       volume = {585},
          eid = {A66},
        pages = {A66},
          doi = {10.1051/0004-6361/201527182},
       adsurl = {https://ui.adsabs.harvard.edu/abs/2016A&A...585A..66B}
}

@ARTICLE{Baran:2019:sdBData,
       author = {{Baran}, A.~S. and {Telting}, J.~H. and {Jeffery}, C.~S. and {{\O}stensen}, R.~H. and {Vos}, J. and {Reed}, M.~D. and {V{\r{A}}­ckovi{\'c}} and {}, M.},
        title = "{K2 observations of the sdBV + dM/bd binaries PHL 457 and EQ Psc}",
      journal = {\mnras},
         year = 2019,
        month = oct,
       volume = {489},
       number = {2},
        pages = {1556-1571},
          doi = {10.1093/mnras/stz2209},
       adsurl = {https://ui.adsabs.harvard.edu/abs/2019MNRAS.489.1556B}
}

@ARTICLE{Sanjayan:2022:sdBData,
       author = {{Sanjayan}, S. and {Baran}, A.~S. and {Ostrowski}, J. and {N{\'e}meth}, P. and {Pelisoli}, I. and {{\O}stensen}, R. and {Kern}, J.~W. and {Reed}, M.~D. and {Sahoo}, S.~K.},
        title = "{Pulsating subdwarf B stars in the oldest open cluster NGC 6791}",
      journal = {\mnras},
         year = 2022,
        month = jan,
       volume = {509},
       number = {1},
        pages = {763-777},
          doi = {10.1093/mnras/stab2985},
archivePrefix = {arXiv},
       eprint = {2111.09873},
 primaryClass = {astro-ph.SR},
       adsurl = {https://ui.adsabs.harvard.edu/abs/2022MNRAS.509..763S}
}

@ARTICLE{Silvotti:2022:sdBData,
       author = {{Silvotti}, Roberto and {N{\'e}meth}, P{\'e}ter and {Telting}, John H. and {Baran}, Andrzej S. and {{\O}stensen}, Roy H. and {Ostrowski}, Jakub and {Sahoo}, Sumanta K. and {Prins}, Saskia},
        title = "{Filling the gap between synchronized and non-synchronized sdBs in short-period sdBV+dM binaries with TESS: TIC 137608661, a new system with a well-defined rotational splitting}",
      journal = {\mnras},
         year = 2022,
        month = apr,
       volume = {511},
       number = {2},
        pages = {2201-2217},
          doi = {10.1093/mnras/stac160},
archivePrefix = {arXiv},
       eprint = {2201.06559},
 primaryClass = {astro-ph.SR},
       adsurl = {https://ui.adsabs.harvard.edu/abs/2022MNRAS.511.2201S}
}

@ARTICLE{Ma:2023:sdBData,
       author = {{Ma}, X.-Y. and {Zong}, W. and {Fu}, J.-N. and {Charpinet}, S. and {Wang}, J. and {Xing}, K.},
        title = "{Amplitude and frequency variations in PG 0101+039 from K2 photometry. A pulsating hot B subdwarf star in an unsynchronized binary system}",
      journal = {\aap},
         year = 2023,
        month = dec,
       volume = {680},
          eid = {A11},
        pages = {A11},
          doi = {10.1051/0004-6361/202347410},
archivePrefix = {arXiv},
       eprint = {2309.11532},
 primaryClass = {astro-ph.SR},
       adsurl = {https://ui.adsabs.harvard.edu/abs/2023A&A...680A..11M}
}

@ARTICLE{Foster:2015:sdBData,
       author = {{Foster}, H.~M. and {Reed}, M.~D. and {Telting}, J.~H. and {{\O}stensen}, R.~H. and {Baran}, A.~S.},
        title = "{The Discovery of Differential Radial Rotation in the Pulsating Subdwarf B Star KIC 3527751}",
      journal = {\apj},
         year = 2015,
        month = jun,
       volume = {805},
       number = {2},
          eid = {94},
        pages = {94},
          doi = {10.1088/0004-637X/805/2/94},
archivePrefix = {arXiv},
       eprint = {1503.02225},
 primaryClass = {astro-ph.SR},
       adsurl = {https://ui.adsabs.harvard.edu/abs/2015ApJ...805...94F}
}

@ARTICLE{Kern:2018:sdBData,
       author = {{Kern}, J.~W. and {Reed}, M.~D. and {Baran}, A.~S. and {Telting}, J.~H. and {{\O}stensen}, R.~H.},
        title = "{Asteroseismic analysis of the pulsating subdwarf B star KIC 11558725: an sdB+WD system with divergent frequency multiplets and mode trapping observed by Kepler}",
      journal = {\mnras},
         year = 2018,
        month = mar,
       volume = {474},
       number = {4},
        pages = {4709-4716},
          doi = {10.1093/mnras/stx2893},
       adsurl = {https://ui.adsabs.harvard.edu/abs/2018MNRAS.474.4709K}
}

@ARTICLE{Su:2024:sdBData,
       author = {{Su}, Wenchao and {Charpinet}, St{\'e}phane and {Latour}, Marilyn and {Zong}, Weikai and {Green}, Elizabeth M. and {Li}, Gang},
        title = "{TIC 441725813: A new bright hybrid hot B subdwarf pulsator with differential core versus envelope rotation}",
      journal = {\aap},
         year = 2024,
        month = oct,
       volume = {690},
          eid = {A36},
        pages = {A36},
          doi = {10.1051/0004-6361/202450020},
archivePrefix = {arXiv},
       eprint = {2407.17887},
 primaryClass = {astro-ph.SR},
       adsurl = {https://ui.adsabs.harvard.edu/abs/2024A&A...690A..36S}
}

@ARTICLE{Geier:2008:sdBData,
       author = {{Geier}, S. and {Nesslinger}, S. and {Heber}, U. and {Randall}, S.~K. and {Edelmann}, H. and {Green}, E.~M.},
        title = "{Tidal synchronisation of the subdwarf B binary PG 0101+039}",
      journal = {\aap},
         year = 2008,
        month = jan,
       volume = {477},
       number = {2},
        pages = {L13-L16},
          doi = {10.1051/0004-6361:20078797},
       adsurl = {https://ui.adsabs.harvard.edu/abs/2008A&A...477L..13G}
}

@ARTICLE{Zong:2018:sdBData,
       author = {{Zong}, Weikai and {Charpinet}, St{\'e}phane and {Fu}, Jian-Ning and {Vauclair}, G{\'e}rard and {Niu}, Jia-Shu and {Su}, Jie},
        title = "{Oscillation Mode Variability in Evolved Compact Pulsators from Kepler Photometry. I. The Hot B Subdwarf Star KIC 3527751}",
      journal = {\apj},
         year = 2018,
        month = feb,
       volume = {853},
       number = {2},
          eid = {98},
        pages = {98},
          doi = {10.3847/1538-4357/aaa548},
       adsurl = {https://ui.adsabs.harvard.edu/abs/2018ApJ...853...98Z}
}

@ARTICLE{Telting:2012:sdBData,
       author = {{Telting}, J.~H. and {{\O}stensen}, R.~H. and {Baran}, A.~S. and {Bloemen}, S. and {Reed}, M.~D. and {Oreiro}, R. and {Farris}, L. and {Ottosen}, T.~A. and {Aerts}, C. and {Kawaler}, S.~D. and {Heber}, U. and {Prins}, S. and {Green}, E.~M. and {Kalomeni}, B. and {O'Toole}, S.~J. and {Mullally}, F. and {Sanderfer}, D.~T. and {Smith}, J.~C. and {Kjeldsen}, H.},
        title = "{Three ways to solve the orbit of KIC 11 558 725: a 10-day beaming sdB+WD binary with a pulsating subdwarf}",
      journal = {\aap},
         year = 2012,
        month = aug,
       volume = {544},
          eid = {A1},
        pages = {A1},
          doi = {10.1051/0004-6361/201219458},
archivePrefix = {arXiv},
       eprint = {1206.3872},
 primaryClass = {astro-ph.SR},
       adsurl = {https://ui.adsabs.harvard.edu/abs/2012A&A...544A...1T}
}

@ARTICLE{Zong:2016:sdBData,
       author = {{Zong}, W. and {Charpinet}, S. and {Vauclair}, G.},
        title = "{Signatures of nonlinear mode interactions in the pulsating hot B subdwarf star KIC 10139564}",
      journal = {\aap},
         year = 2016,
        month = oct,
       volume = {594},
          eid = {A46},
        pages = {A46},
          doi = {10.1051/0004-6361/201629132},
archivePrefix = {arXiv},
       eprint = {1607.06621},
 primaryClass = {astro-ph.SR},
       adsurl = {https://ui.adsabs.harvard.edu/abs/2016A&A...594A..46Z}
}

@ARTICLE{Schaffenroth:2014:sdBData,
       author = {{Schaffenroth}, V. and {Classen}, L. and {Nagel}, K. and {Geier}, S. and {Koen}, C. and {Heber}, U. and {Edelmann}, H.},
        title = "{Two candidate brown dwarf companions around core helium-burning stars}",
      journal = {\aap},
         year = 2014,
        month = oct,
       volume = {570},
          eid = {A70},
        pages = {A70},
          doi = {10.1051/0004-6361/201424616},
archivePrefix = {arXiv},
       eprint = {1409.4357},
 primaryClass = {astro-ph.SR},
       adsurl = {https://ui.adsabs.harvard.edu/abs/2014A&A...570A..70S}
}

@ARTICLE{Reed:2025:sdBData,
       author = {{Reed}, M.~D. and {Baran}, A.~S. and {Telting}, J.~H. and {{\O}stensen}, R.~H. and {Jeffery}, C.~S. and {Gaibor}, Y.},
        title = "{K2 observations of five pulsating subdwarf-B stars with white dwarf companions}",
      journal = {\mnras},
         year = 2025,
        month = jul,
       volume = {540},
       number = {4},
        pages = {3725-3752},
          doi = {10.1093/mnras/staf917},
archivePrefix = {arXiv},
       eprint = {2506.05033},
 primaryClass = {astro-ph.SR},
       adsurl = {https://ui.adsabs.harvard.edu/abs/2025MNRAS.540.3725R}
}
\bibliographystyle{mnras}

\begin{landscape}
\begin{table}
\caption{Observed and model properties and upper limits on the magnetic field of \NSDB{} sdBs}\label{tab:TAB_field_constraints}
\centering
\begin{tabular}{l cccc cc cccc cccc}
\toprule
 & \multicolumn{6}{c}{Observed Properties} & \multicolumn{4}{c}{Most Conservative $\ell=1$ Model} & \multicolumn{4}{c}{Most Conservative $\ell=2$ Model} \\
\cmidrule(lr){2-7}\cmidrule(lr){8-11}\cmidrule(lr){12-15}
Name & $T_{\mathrm{eff}}$ & $\log g$ & $\log(n_{\mathrm{He}}/n_{\mathrm{H}})$ & companion & $\nu^{\ell=1}_{\mathrm{min}}$ & $\nu^{\ell=2}_{\mathrm{min}}$ & $M$ & $M_{\mathrm{H}}$ & $Y_c$ & $B_{r,\mathrm{supp}}$ & $M$ & $M_{\mathrm{H}}$ & $Y_c$ & $B_{r,\mathrm{supp}}$\\
 & K & $\log(\mathrm{cm}/\mathrm{s}^2)$ & & & $\mu$Hz & $\mu$Hz & $M_\odot$ & $M_\odot$ & \% & G & $M_\odot$ & $M_\odot$ & \% & G\\
\midrule
EQ Psc$^{1}$ & $28320\pm50$ & $5.632\pm0.015$ & $-2.80\pm0.06$ & dM$^{1}$ & 197.14 & -- & 0.47 & $4.5\,10^{-4}$ & 55.7 & \underline{10k} & -- & -- & -- & --\\
Feige 46$^{\new{*},2}$ & $36120\pm230$ & $5.93\pm0.04$ & $-0.32\pm0.03$ & -- & 294.06 & -- & 0.43 & $5.0\,10^{-4}$ & 97.9 & \underline{30k} & -- & -- & -- & --\\
HZ Cnc$^{3}$ & $27557\pm85$ & $5.430\pm0.014$ & $-2.811\pm0.025$ & WD$^{4}$ & 112.14 & 226.73 & 0.47 & $8.5\,10^{-4}$ & 12.0 & \underline{2k} & 0.47 & $8.5\,10^{-4}$ & 12.7 & 8k\\
KIC 2437937$^{5}$ & $24860\pm270$ & $5.348\pm0.052$ & $-2.46\pm0.12$ & -- & 91.33 & 239.54 & 0.48 & $1.9\,10^{-3}$ & 16.5 & \underline{2k} & 0.48 & $1.9\,10^{-3}$ & 16.5 & 30k\\
KIC 2438324$^{5}$ & $25290\pm300$ & $5.510\pm0.043$ & $-2.62\pm0.11$ & dM$^{6}$ & 116.41 & 339.33 & 0.48 & $1.7\,10^{-3}$ & 59.9 & \underline{2k} & 0.48 & $1.7\,10^{-3}$ & 59.9 & 80k\\
KIC 2569576$^{5}$ & $23540\pm210$ & $5.311\pm0.035$ & $-2.73\pm0.23$ & -- & 122.04 & 127.47 & 0.48 & $2.8\,10^{-3}$ & 41.5 & 10k & 0.38 & $2.1\,10^{-3}$ & 98.0 & \underline{1k}\\
KIC 2697388$^{7}$ & $23900\pm300^{8}$ & $5.32\pm0.03^{8}$ & $-2.9\pm0.1^{8}$ & -- & 89.11 & 111.48 & 0.48 & $2.4\,10^{-3}$ & 37.5 & 1k & 0.38 & $2.1\,10^{-3}$ & 98.0 & \underline{500}\\
KIC 2991403$^{7,9}$ & $27300\pm200^{8}$ & $5.43\pm0.03^{8}$ & $-2.6\pm0.1^{8}$ & dM$^{9}$ & 156.61 & 157.11 & 0.47 & $1.0\,10^{-3}$ & 11.6 & 20k & 0.47 & $1.0\,10^{-3}$ & 11.6 & \underline{1k}\\
KIC 3527751$^{7,10,11}$ & $27818\pm163$ & $5.35\pm0.03$ & $-2.99\pm0.04$ & -- & 87.85 & 91.70 & 0.44 & $2.1\,10^{-3}$ & 97.9 & 1k & 0.44 & $2.1\,10^{-3}$ & 97.9 & \underline{200}\\
KIC 5807616$^{7}$ & $27100\pm200^{8}$ & $5.51\pm0.02^{8}$ & $-2.9\pm0.1^{8}$ & -- & 109.50 & 225.90 & 0.44 & $1.1\,10^{-3}$ & 97.9 & \underline{900} & 0.47 & $8.0\,10^{-4}$ & 35.4 & 5k\\
KIC 7664467$^{7,12}$ & $27440\pm120$ & $5.382\pm0.019$ & $-2.57\pm0.05$ & WD$^{12}$ & 91.81 & 208.39 & 0.48 & $1.5\,10^{-3}$ & 2.7 & \underline{800} & 0.48 & $1.5\,10^{-3}$ & 2.7 & 7k\\
KIC 7668647$^{7,13}$ & $27700\pm300$ & $5.50\pm0.03$ & $-2.65\pm0.03$ & WD$^{13}$ & 109.96 & 109.75 & 0.47 & $7.5\,10^{-4}$ & 19.6 & 1k & 0.44 & $1.1\,10^{-3}$ & 97.9 & \underline{80}\\
KIC 8302197$^{7}$ & $26400\pm300^{14}$ & $5.32\pm0.04^{14}$ & $-2.7\pm0.1^{14}$ & -- & 126.30 & -- & 0.48 & $2.8\,10^{-3}$ & 1.0 & \underline{10k} & -- & -- & -- & --\\
KIC 9472174$^{15,7}$ & $29564\pm106^{8}$ & $5.425\pm0.009^{8}$ & $-2.4\pm0.1^{8}$ & dM$^{15}$ & 57.96 & -- & 0.47 & $1.4\,10^{-3}$ & 1.0 & \underline{20} & -- & -- & -- & --\\
KIC 10001893$^{7,16}$ & $26700\pm300$ & $5.30\pm0.04$ & $-2.9\pm0.1$ & -- & 102.17 & 181.94 & 0.48 & $2.7\,10^{-3}$ & 1.1 & \underline{3k} & 0.48 & $2.7\,10^{-3}$ & 1.1 & 6k\\
KIC 10139564$^{17,18}$ & $31859\pm126$ & $5.673\pm0.026$ & $-2.201\pm0.036$ & -- & 315.58 & 345.25 & 0.45 & $4.1\,10^{-3}$ & 96.6 & 200k & 0.45 & $4.1\,10^{-3}$ & 96.6 & \underline{30k}\\
KIC 10553698$^{7}$ & $27600\pm400^{14}$ & $5.33\pm0.05^{14}$ & $-2.9\pm0.2^{14}$ & WD$^{19}$ & 140.51 & 104.29 & 0.44 & $2.1\,10^{-3}$ & 97.9 & 20k & 0.44 & $2.1\,10^{-3}$ & 97.9 & \underline{400}\\
KIC 10670103$^{7}$ & $20900\pm300^{8}$ & $5.11\pm0.04^{8}$ & $-2.2\pm0.3^{8}$ & -- & 61.39 & 138.10 & 0.38 & $3.0\,10^{-3}$ & 98.0 & \underline{1k} & 0.48 & $5.1\,10^{-3}$ & 38.2 & 7k\\
KIC 11179657$^{7,9}$ & $26000\pm800$ & $5.14\pm0.13$ & $-2.1\pm0.1$ & dM$^{9}$ & 185.72 & 99.56 & 0.38 & $3.1\,10^{-3}$ & 98.0 & 500k & 0.38 & $3.1\,10^{-3}$ & 98.0 & \underline{2k}\\
KIC 11558725$^{7,10,20}$ & $27910\pm210$ & $5.410\pm0.015$ & $-3.116\pm0.018$ & WD$^{10}$ & 66.56 & 83.28 & 0.47 & $1.1\,10^{-3}$ & 4.3 & 60 & 0.47 & $1.1\,10^{-3}$ & 4.3 & \underline{20}\\
KPD 0629--0016$^{7}$ & $27781\pm270^{21}$ & $5.527\pm0.041^{21}$ & $-2.746\pm0.118^{21}$ & WD$^{22}$ & 95.09 & 130.88 & 0.44 & $1.1\,10^{-3}$ & 97.9 & 600 & 0.44 & $1.1\,10^{-3}$ & 97.9 & \underline{300}\\
LT Cnc$^{3}$ & $26032\pm83$ & $5.275\pm0.011$ & $-2.801\pm0.041$ & WD$^{4}$ & 111.89 & 183.08 & 0.48 & $2.1\,10^{-3}$ & 3.0 & \underline{5k} & 0.48 & $2.1\,10^{-3}$ & 3.0 & 7k\\
PB 6373$^{3}$ & $26968\pm125$ & $5.404\pm0.018$ & $-3.01\pm0.04$ & WD$^{23}$ & 108.30 & 193.47 & 0.47 & $1.0\,10^{-3}$ & 12.3 & \underline{2k} & 0.47 & $1.0\,10^{-3}$ & 12.3 & 4k\\
PG 0101+039$^{24,3}$ & $27145\pm111$ & $5.53\pm0.07$ & $-2.750\pm0.032$ & WD$^{25}$ & 144.03 & 111.64 & 0.47 & $9.1\,10^{-3}$ & 10.1 & 9k & 0.44 & $1.1\,10^{-3}$ & 97.9 & \underline{90}\\
PG 0902+124$^{3}$ & $27287\pm77$ & $5.46\pm0.01$ & $-2.726\pm0.034$ & WD$^{26}$ & 101.25 & 275.67 & 0.47 & $7.5\,10^{-4}$ & 26.6 & \underline{600} & 0.47 & $7.5\,10^{-4}$ & 26.6 & 20k\\
PHL 417$^{\new{*},27}$ & $35640\pm280$ & $5.73\pm0.01$ & $-0.49\pm0.02$ & -- & 146.29 & -- & 0.43 & $3.0\,10^{-4}$ & 97.9 & \underline{1k} & -- & -- & -- & --\\
PHL 457$^{1}$ & $26690\pm60$ & $5.312\pm0.008$ & $-2.487\pm0.022$ & BD$^{28}$ & 152.47 & -- & 0.48 & $1.7\,10^{-3}$ & 3.1 & \underline{20k} & -- & -- & -- & --\\
TIC 137608661$^{29}$ & $27660\pm64$ & $5.394\pm0.009$ & $-2.945\pm0.012$ & dM$^{29}$ & 155.68 & 276.54 & 0.47 & $1.1\,10^{-3}$ & 4.9 & \underline{10k} & 0.47 & $1.1\,10^{-3}$ & 4.9 & 30k\\
TIC 441725813$^{30}$ & $27827\pm177$ & $5.463\pm0.028$ & $-2.826\pm0.128$ & WD?$^{30}$ & 72.77 & 153.41 & 0.47 & $8.5\,10^{-4}$ & 9.7 & \underline{90} & 0.47 & $8.5\,10^{-4}$ & 9.7 & 600\\
\new{V366 Aqr}$^{\new{*},31,32}$ & $35150\pm111^{33}$ & $5.88\pm0.02^{33}$ & $-0.62\pm0.01^{33}$ & -- & 196.71 & -- & 0.37 & $7.0\,10^{-4}$ & 98.0 & \underline{400} & -- & -- & -- & --\\
\bottomrule
\end{tabular}
\\
\raggedright For \NSDB{} \new{hot subdwarfs}: observed spectroscopic parameters $T_{\mathrm{eff}}$, $\log g$, and $\log(n_{\mathrm{He}}/n_{\mathrm{H}})$, asteroseismic upper limits on the radial component of the magnetic field $B^{\ell=1}_{r,\mathrm{supp}}$ and $B^{\ell=2}_{r,\mathrm{supp}}$ derived from the lowest-frequency $\ell=1$ and $\ell=2$ g modes with respective frequencies $\nu^{\ell=1}_{\mathrm{min}}$ and $\nu^{\ell=2}_{\mathrm{min}}$, and the total masses $M$, hydrogen masses $M_{\mathrm{H}}$, and central helium abundances $Y_c$ of the models used to derive them. For each star, we underline the most stringent upper limit on the magnetic field. \new{Systems denoted by $*$ have unusual high effective temperatures that may indicate peculiar chemical enrichment that are not captured in our models, hence their field estimates should be treated with cautions (see discussions in \ref{sec:bounding_fields}). References:} 1:~\citet{Baran:2019:sdBData} 2:~\citet{Latour:2019:sdBData} 3:~\citet{Reed:2025:sdBData} 4:~\citet{Morales-Rueda:2003:sdBData} 5:~\citet{Sanjayan:2022:sdBData} 6:~\citet{Pablo:2011:sdBData} 7:~\citet{Reed:2011:sdBData} 8:~\citet{Oestensen:2010A:sdBData} 9:~\citet{Pablo:2012:sdBData} 10:~\citet{Telting:2012:sdBData} 11:~\citet{Foster:2015:sdBData} 12:~\citet{Baran:2016:sdBData} 13:~\citet{Telting:2014:sdBData} 14:~\citet{Baran:2011:sdBData} 15:~\citet{Oestensen:2010:sdBData} 16:~\citet{Uzundag:2017:sdBData} 17:~\citet{Baran:2012:sdBData} 18:~\citet{Zong:2016:sdBData} 19:~\citet{Oestensen:2014:sdBData} 20:~\citet{Kern:2018:sdBData} 21:~\citet{Koen:2007:sdBData} 22:~\citet{Schaffenroth:2023:sdBData} 23:~\citet{Kupfer:2015:sdBData} 24:~\citet{Ma:2023:sdBData} 25:~\citet{Geier:2008:sdBData} 26:~\citet{Moran:1999:sdBData} 27:~\citet{Oestensen:2020:sdBData} 28:~\citet{Schaffenroth:2014:sdBData} 29:~\citet{Silvotti:2022:sdBData} 30:~\citet{Su:2024:sdBData} 31:~\citet{Ahmad:2005:sdBData} 32:~\citet{Green:2011:sdBData} 33:~\citet{Randall:2015:sdBData}
\end{table}
\end{landscape}

\end{CJK*}
\end{document}